\documentclass[fleqn,usenatbib]{mnras}

\usepackage{newtxtext,newtxmath,lscape}

\usepackage[T1]{fontenc}
\usepackage{ae,aecompl}

\usepackage{graphicx}	
\usepackage{amsmath}	
\usepackage{multicol}

\title[Dynamical friction modelling]{A calibrated model for N-body dynamical friction acting on supermassive black holes}
\author[A. Genina et al.]{Anna Genina$^{1}$\thanks{E-mail: agenina@mpa-garching.mpg.de (AG)},  Volker Springel$^{1}$, Antti Rantala$^{1}$  \vspace*{0.1cm}\\%
$^{1}$Max-Planck-Institut für Astrophysik, Karl-Schwarzschild-Str. 1, D-85748, Garching, Germany\\%
}

\date{Accepted XXX. Received YYY; in original form ZZZ}

\pubyear{2023}

\begin{document}
\label{firstpage}
\pagerange{\pageref{firstpage}--\pageref{lastpage}}
\maketitle

\begin{abstract}
Black holes are believed to be crucial in regulating star formation in massive galaxies, which makes it essential to faithfully represent the physics of these objects in cosmological hydrodynamics simulations. Limited spatial and mass resolution and the associated discreteness noise make following the dynamics of black holes especially challenging. In particular, dynamical friction, which is responsible for driving massive black holes towards the centres of galaxies, cannot be accurately modelled with softened $N$-body interactions. A number of subgrid models have been proposed to mimic dynamical friction or directly include its full effects in simulations. Each of these methods has its individual benefits and shortcomings, while all suffer from a common issue of being unable to represent black holes with masses below a few times the simulated dark matter particle mass. In this paper, we propose a correction for unresolved dynamical friction, which has been calibrated on simulations run with the code {\small KETJU}, in which gravitational interactions of black holes are not softened. We demonstrate that our correction is able to sink black holes with masses greater than the dark matter particle mass at the correct rate. We show that the impact of stochasticity is significant for low-mass black holes ($M_{\rm BH} \leq 5 M_{\rm DM}$) and propose a correction for stochastic heating. Combined, this approach is applicable to next generation cosmological hydrodynamics simulations that jointly track galaxy and black hole growth with realistic black hole orbits.
\end{abstract}




\section{Introduction}
\label{intro}

Most galaxies are observed to harbour supermassive black holes (BHs) close to their centres \citep[see][for a review]{Kormendy2013}. These BHs are not only fascinating objects in their own right, they are also recognized to play a decisive role in galaxy formation and evolution \citep{Richstone1998}. During periods of rapid growth through gas accretion, supermassive BHs can liberate enormous amounts of energy, shine as quasars, and potentially influence their host galaxies through this energy output \citep{Silk1998, DiMatteo2005}. But even in low states of accretion, the associated radio activity may arrest cooling flows in early-type galaxies and explain why their star formation is quenched \citep{Croton2006}. Modelling massive galaxy formation is thus not possible without addressing the evolution of supermassive black holes at the same time. In addition, supermassive BHs are also of interest as sources of powerful gravitational waves, emitted when they merge during the hierarchical galaxy formation process \citep[e.g.][]{Begelman1980,Ebisuzaki1991,Jaffe2003}.

Modern models of galaxy formation therefore aim to jointly study the evolution of galaxies and supermassive black holes. This is achieved either with simplified semi-analytic treatments \citep[e.g.][]{santa_cruz, Kauffmann2000, cole_galform} or with hydrodynamical simulations \citep[e.g.][]{Weinberger2017}, where the latter approach has been pioneered nearly two decades ago \citep{sdmh,DiMatteo2005} and matured to a point where current cosmological hydrodynamical simulation studies such as Illustris \citep{Vogelsberger2014}, Eagle \citep{Schaye2015}, MassiveBlack \citep{Khandai2015}, HorizonAGN \citep{Dubois2016}, Magneticum \citep{Dolag2016}, IllustrisTNG \citep{Pillepich2018}, or Simba \citep{Dave2019} are able to successfully reproduce many galaxy and black hole scaling relations and properties, including the quenching transition of massive galaxies that is related to their embedded black holes \citep{Nelson2018}. 

These galaxy formation models involve, however, numerous coarse assumptions to make the problem at least approximately tractable. These simplifications first of all relate to the accretion and feedback processes that occur during the growth of black holes, which cannot be simulated in an ab initio fashion in today's cosmological simulations due to the associated huge dynamic range of spatial and temporal scales. Instead, the black holes are represented as sink particles for which the accretion, feedback and merger physics is prescribed in terms of a subgrid model. But even the basic tracking of the spatial coordinates of these sink particles, which corresponds to the orbits of the black holes, is fraught with substantial difficulty in simulations of whole galaxies. This is because the typical mass resolution of such simulations for the dark matter component is rarely better than $\sim 10^5\,{\rm M}_\odot$ (TNG50; \citealt{tng50}). In fact, it is often orders of magnitude worse than that, especially when large cosmological volumes are studied, meaning that the physical masses of many supermassive black holes are smaller or comparable to the mass of dark matter simulation particles ($\sim 10^4 - 10^5\,{\rm M}_{\odot}$ ). This immediately implies that fiducial BH sink particles will experience substantial numerical heating effects in N-body models due to discreteness \citep{ludlow}.

As a consequence of these numerical influences, the orbits of numerically represented supermassive black holes can easily become unphysical. In particular, the process of dynamical friction, which drives massive black holes to the centres of their host galaxies \citep{chandrasekhar1943, binney1977, colpi_1999}, cannot be fully resolved in modern cosmological simulations. This arises both due to the coarseness in the background particle representation and the use of gravitational softening, which limits the effects of gravitational encounters of BHs with the surrounding matter. As a result, simulated BHs may not sink to the centres of galaxies on appropriate timescales, or get spuriously ejected from galaxy centres. Since the densest regions are normally found at the very centre of a galaxy, this will invariably also modify the growth rate of BHs, which in turn impacts the strength of feedback processes as well as their coupling efficiency to the host galaxy \citep{callegari}. It will also affect how quickly BHs come together after a galaxy merger, form hard BH binaries, and eventually coalesce. Clearly, a failure to properly account for the orbits of supermassive BHs can be a significant stumbling block for a realistic treatment of the co-evolution of galaxies and black holes.

A number of methods have therefore been proposed in the literature, aimed at correcting numerical effects that lead to supermassive black holes wandering away from the centres of their galaxies. These numerical effects are, however, distinct from physical processes that lead to similar outcomes, for instance due to black hole merger recoils from asymmetric gravitational wave emission, or slingshot kicks in 3-body black hole interactions. The prevalence of these physical processes is not yet fully understood. Their outcomes are believed to be close to stochastic \citep{rawlings,partmann}, while their modeling would require extremely high temporal, spatial and mass resolution that is currently not achievable in simulations of large cosmological volumes. We therefore focus in this study exclusively on the problem of black hole sinking due to dynamical friction. In the following, we first motivate our work by outlining some of the existing techniques and comment on their successes and shortcomings. 

\subsection{Black hole repositioning methods}

Repositioning methods generally do not intend to model the orbital evolution of black holes, but rather aim to keep black holes near the minimum of the gravitational potential by updating their position to the location of the most bound particle within some search kernel around the black hole at each timestep \citep{sdmh2,booth_schaye}. Differences in implementations mainly arise in how the minimum-potential particle is selected, in particular the type of the particle: gas (IllustrisTNG; \citealt{Weinberger2017}), or all particle types (EAGLE; \citealt{eagle_schaye}). Additional constraints are sometimes invoked based on the velocity of the black hole particle compared to the local gas bulk or sound speed, to avoid black holes being dragged by gas particles with low potential, but high kinetic energy \citep{bahe_repos}. 

Repositioning methods tend to sink black holes to the centre of the galactic potential too fast; however, despite this aggressive treatment, this method has been shown to suffer from other unphysical behaviours such as removal of black holes from backsplash galaxies \citep{borrow}, or strongly enhanced merging of black holes and their seeds at high redshift \citep{ma_seeds, chen}. Some more recent works \citep{rhapsody-c} have introduced new repositioning models motivated by gradient descent methods that, to an extent, mimic the orbital decay of a black hole. The sinking time of the black hole can also be increased or decreased using a weight function that determines how aggressively the black hole is drifted in the direction of the potential gradient, and by limiting its velocity using a Courant-like criterion. While delaying the sinking time of the black holes compared to standard repositioning methods, the gradient descent models still cannot accurately reproduce the sinking rate of individual black holes.

\subsection{Compensating for unresolved dynamical friction}

A number of works based their subgrid dynamical friction prescription on the 
Chandrasekhar formula \citep{chandrasekhar1943}, describing the deceleration that a massive body experiences when traversing a ``sea'' of isotropically distributed low-mass particles:
\begin{equation}
    \mathbf{a_{\rm df}} = -16 \pi^2 \ln \Lambda \,G^2 M_{\rm BH} \frac{\int_{0}^v f(u) u^2 {\rm d}u}{v^3} \mathbf{v},
    \label{chandrasekhar_formula}
\end{equation}
where $G$ is the gravitational constant, $f(u)$ is the distribution function of background particles (where often the Maxwellian distribution is assumed), $M_{\rm BH}$ is the black hole mass, $v$ is the black hole velocity and $\ln \Lambda = \ln \frac{b_{\rm max}}{b_{\rm min}}$ is the Coulomb logarithm. The minimum impact parameter, $b_{\rm min}$, is the impact parameter for a 90-degree deflection. The maximum impact parameter, $b_{\rm max}$, is not well defined but, in principle, describes the largest impact parameter where the gravitational scattering of a background particle still contributes to dynamical friction.

The work of \citet{tremmel} introduced an estimator aimed at compensating for unresolved dynamical friction below the gravitational softening scale,
\begin{equation}
\mathbf{a}_{\rm df,T} = -4 \pi G^2 M_{\rm BH} \rho(v<v_{\rm BH}) \ln \Lambda_{\rm T} \frac{{\mathbf{v}}_{\rm BH}}{v^3_{\rm BH}},
\label{equation_tremmel}
\end{equation}
where $\rho(v<v_{\rm BH})$ is the density of particles moving slower than the black hole, measured in a kernel covering 64 nearest collisionless particles. 
The Coulomb logarithm is here defined as:
\begin{equation}
\Lambda_{\rm T} =  \frac{b_{\rm max,unres}}{(GM_{\rm BH})/v^2_{\rm BH}},
\end{equation}
where $v_{\rm BH}$ is the velocity of the black hole relative to the surrounding medium, and the maximum unresolved impact parameter, $b_{\rm max, unres}$, is taken to be the gravitational softening, $\epsilon_g$. Such a correction has been employed in the Romulus cosmological simulations suite to study supermassive BH dynamics and populations of wandering black holes \citep{romulus1, romulus2, ricarte_wbh}.

As pointed out by \citet{chen}, the kernel over the nearest 64-100 particles can be noisy and the choice of $\epsilon_g$ as the maximum unresolved impact parameter is not necessarily justified, given that the forces are softened compared to a Newtonian case beyond the scale of gravitational softening in many simulations (but dependent on how softening is implemented). It is also worth noting that, for particularly small kernel sizes, there may be no particles moving slower than the black hole. \citet{chen} instead propose the formula:
\begin{equation}
\mathbf{a}_{\rm df,C} = -4 \pi G^2 M_{\rm BH} \rho_{\rm sph} \mathcal{F}\left(x \right) \ln \Lambda_{\rm C}  \frac{{\mathbf{v}}_{\rm BH}}{v^3_{\rm BH}},
\label{equation_chen}
\end{equation}
where $\rho_{\rm sph}$ is the density of collisionless particles measured within the Smooth Particle Hydrodynamics (SPH) kernel employed in their simulations, and $\mathcal{F}(x)$ describes a Maxwellian local velocity distribution:
\begin{equation}
\mathcal{F}(x) = {\rm erf}(x) - \frac{2x}{\sqrt{\pi}} e^{-x^2} , \;\;\;\; x = \frac{v_{\rm BH}}{\sqrt{2}\sigma_v},
\end{equation}
where $\sigma_v$ is the one-dimensional velocity dispersion of particles in the SPH kernel, and $v_{\rm BH}$ denotes the velocity of the black hole relative to the surrounding medium. The Coulomb logarithm is again defined as:
\begin{equation}
\Lambda_{\rm C} = \frac{b_{\rm max, unres}}{(GM_{\rm BH})/v^2_{\rm BH}} = \frac{\eta \epsilon_g}{(GM_{\rm BH})/v^2_{\rm BH}}, 
\end{equation}
where $\eta$ is a multiple of the gravitational softening $\epsilon_g$ below which dynamical friction is not resolved. In the simulations of \citet{chen}, this was found to be close to $\eta = 6$.

This method  of \citet{chen} has been shown to be less sensitive to noise (due to the use of the SPH kernel in computing average background particle properties) and allows one to take the scale below which the dynamical friction is unresolved as a free parameter of the simulation. The method has been employed in the cosmological simulation suite ASTRID \citep{astrid, astrid_wbh}.

The method of \citet{pfister} is similar to those described above, but takes into account the contribution of fast-moving particles to dynamical friction, which may be important in enhancing dynamical friction in cases where the density distribution of the background particles becomes shallow \citep{antonini_merritt, dosopoulou_antonini}, such as in central density cores \citep{deBlok}. However, we also note that the ``standard'' Chandrasekhar treatment is known to fail in reproducing the effect of stalling in cored density distributions \citep{read,kaur,banik}.

\subsection{Dynamical friction from $N$-body particle distributions}

The work of \citet{ma_hopkins} points out a number of shortcomings in the application of the Chandrasekhar formula to model dynamical friction in simulations, namely the assumptions of an isotropic and homogeneous medium, ambiguities in defining the Coulomb logarithm, as well as difficulties in determining the local velocity distribution. The authors instead present a derivation of a discrete dynamical friction formula, allowing the full dynamical friction force to be computed self-consistently in $N$-body simulations. \citet{ma_hopkins} also provide an expression for the perpendicular component of dynamical friction which arises due to the presence of a local density gradient, in addition to that parallel to the black hole velocity vector. However, the contribution of the transverse component is expected to be relatively minor (see also \citealt{just_penarrubia}). 

Using their method, the dynamical friction force can be computed straightforwardly with direct summation, or with tree methods by treating a distant node as a single particle. Manifest momentum conservation can also be introduced by applying an opposite force to each particle that contributes dynamical friction to the black hole -- a feature that cannot be self-consistently implemented alongside models of \citet{tremmel} and \citealt{chen} (as well as the model presented in this work, see Appendix C). Nevertheless, for relatively well-resolved cases, such an approach, in effect, double-counts the dynamical friction because a fraction of the force is already resolved in the simulation and taken into account in the force computation. \citet{ma_hopkins} acknowledge this shortcoming and point out that this can be corrected in the future by some mass-resolution-dependent weight function. 

For low-resolution cases, the noisy nature of the forces also translates into a noisy dynamical friction estimate. Moreover, as the approach of \citet{ma_hopkins} takes into account gravitational softening in their computation of dynamical friction, it is implied that the impact parameters below the softening length scale are still unaccounted for. Ultimately, when defining a dynamical friction model, one needs to decide whether they aim to model the orbits of black holes (independent of mass) in the {\it softened} $N$-body potential or in the fiducial {\it infinitely high-resolution} $N$-body potential that the simulation is aiming to sample. In this work, we take the latter approach and aim for convergence in black hole sinking timescales across resolution levels.

The recent work of \citet{damiano} introduces a discrete estimator for \textit{unresolved} dynamical friction, where each particle lying within the softening length of the black hole contributes to an approximation of the local distribution function and exerts a force on the black hole in the direction defined by the vector ${\bf v_{\rm BH} - v_{\rm m}}$, with $v_{\rm m}$ the velocity of the particle. The maximum impact parameter in the Coulomb logarithm, as in \citet{tremmel}, is assumed to be the softening length, while the minimum impact parameter for each encounter is defined by the relative velocity of the two interacting particles, $b_{\rm min} = \frac{G(M+m)}{(v_{\rm BH}-v_{\rm m})^2}$. This method is thus able to treat interactions with multiple populations of background particles, which can differ in their mass and velocity distribution, consistently, without grouping them into a background ``sea'' of particles as in the standard approach of \citet{chandrasekhar1943}. However, a clear disadvantage of such a treatment is that its effectiveness relies on the presence of particles within the softening length. Issues may occur in the outskirts of halos, where not many particles are present, or even closer to their centres, if the gravitational softening is set too low for the halo mass resolution. 

\subsection{Partial or no added dynamical friction}

Finally, we note that in simulations like {\sc Horizon-AGN} no additional dynamical friction force due to collisionless particles is applied, instead relying on the dynamical friction already present in the simulations when the black hole mass is higher than that of the dark matter particles \citep{horizon1, horizon2} and an additional force due to gas drag \citep{ostriker_gas_drag,chapon}. In {\sc magneticum} simulations, the black holes are seeded at the location of the star particles with highest binding energies, ensuring stability of the black hole position against mergers and local gas flows, while the dynamical friction force following the Chandrasekhar formula is applied over all impact parameters out to the halo size for as long as the black holes are less massive than the dark matter particle mass \citep{hirschmann,steinborn_magneticum}.

\subsection{This work}

Perhaps the biggest shortcoming of the available subgrid dynamical friction models is a lack of model calibration on high-resolution simulations with unsoftened black hole interactions, in which dynamical friction is fully resolved. In this work, we achieve this through comparison to simulations run with the code {\small KETJU}. We chose to base our dynamical friction model on the previous works of \citet{tremmel} and \citet{chen}, as these methods, in principle, only make assumptions about the effect of dynamical friction in the direct vicinity of the black hole, modified by gravitational softening, while dynamical friction due to distant particles is already accounted for in the simulation \citep{spinnato}. We do not consider the effect of gas drag in this work, which can be important for black hole dynamics, particularly at high redshift \citep{ostriker_gas_drag, chapon, escala, chen}. We note, however, that our findings regarding the effect of gravitational softening on dynamical friction from collisionless particles in simulations may be applicable to the treatment of gas drag in the future.

This paper is structured as follows. In Section~\ref{methods} we introduce our numerical set-up and the simulations with {\small KETJU}. In Section~\ref{trem_chen_sec}~we evaluate how the methods of \citet{tremmel} and \citet{chen} behave in this set-up. In Section~\ref{my_model_sec}, we outline our calibration procedure to determine the optimal set of parameters of our dynamical friction model. In Section~\ref{summary_sec}, we summarise the required steps for implementation of our model and apply it to a number of test cases. Finally, in Section~\ref{conclusion_sec}, we present our conclusions.










\section{Methods}
\label{methods}

In this section, we consider how well dynamical friction is resolved in $N$-body simulations. To be able to quantify this, we must first define what the true rate of black hole orbital decay is in a particular set-up. For the latter, we will consider both the analytical prediction of the Chandrasekhar formula as well as the result of high resolution simulations with {\small KETJU}, the details of which we outline below.

\begin{table}
\centering
\caption{The black hole to dark matter mass ratio, the number of particles used, the number of particles within the original orbital radius of the black hole and the employed gravitational softening for our primary set of simulations carried out with {\small GADGET-4}. In bold, we show the resolution of the simulation carried out with {\small KETJU}.  }
\begin{tabular}{c|r|r|c}
\hline
$M_{\rm BH}/M_{\rm DM}$ & $N_{\rm DM}$& $N_{\rm DM} (<20{\rm kpc})$ & $\epsilon_g$ [kpc] \\
\hline
1      & $10^4$     & $4 \times 10^2$   & 14.00   \\
5      & $5 \times 10^4$ & $2 \times 10^3$ & 6.27 \\
10     & $10^5$    &  $4 \times 10^3$   & 4.43 \\
50     & $5 \times 10^5$& $2 \times 10^4$ & 1.98 \\
100    & $10^6$  &  $4 \times 10^4$     & 1.40  \\
500    & $5 \times 10^6$& $2 \times 10^5$ & 0.63 \\
1000   & $10^7$  &  $4 \times 10^5$     & 0.44 \\
{\bf 5000}   & $\mathbf{5 \times 10^7}$ &  $\mathbf{2 \times 10^6}$      & {\bf 0.20} \\
10000  & $10^{8}$ & $4 \times 10^6$   & 0.14 \\
\hline
\end{tabular}
\label{table1}
\end{table}

\begin{figure*}
    \centering
    
    \includegraphics[width = 2\columnwidth]{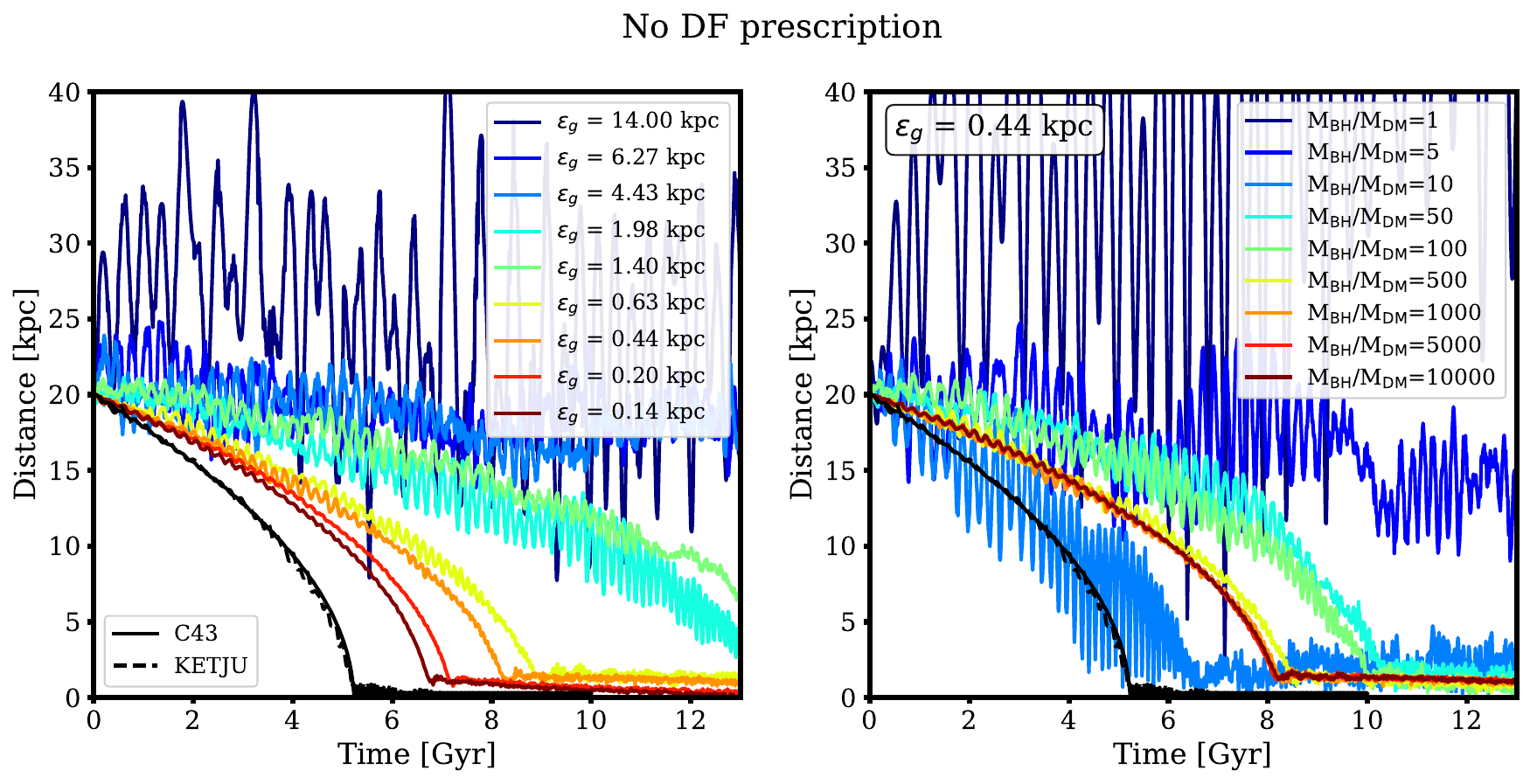}\\

   \caption{\textit{Left:} Distance of the black hole from the centre of the halo, as determined by the shrinking spheres algorithm \citep{power}, as a function of time, for simulations with {\small GADGET-4} that do not include a dynamical friction prescription. Blue to red lines show simulations varying from lowest to highest resolution, with the softening chosen optimally for each halo mass resolution. The black solid line is the analytical prediction using the Chandrasekhar formula (C43), and the black dashed line is the result of the run with {\small KETJU}. \textit{Right:} as the left figure, but using a fixed softening of 0.44~kpc for all mass resolutions. }
    \label{fig1}  
\end{figure*}

\subsection{Simulations with {\small GADGET-4}}

We set up a dark matter halo of mass $10^{13}\, {\rm M}_{\odot}$ following a Hernquist density profile \citep{hernquist}, using the {\small MAKENEWDISK} code \citep{sdmh2}. A black hole with mass $M_{\rm BH} = 10^9\,{\rm M}_{\odot}$ is set on a circular orbit, starting at 20~kpc, corresponding to $1/4$ of the scale radius of the Hernquist halo. We note that this set-up by no means represents a typical cosmological scenario, but we aimed to chose a test case such that {\it i)} a black hole starting off on a circular orbit sinks within the age of the Universe and {\it ii)} the sinking occurs over a sufficiently extended period of time such that we have a good probe of the black hole dynamics and the differences from predictions of high-resolution simulations with {\small KETJU} become clearly identifiable. The black hole must be massive enough to sink to the centre, but not too massive such that it causes significant motion in the centre of mass of the halo.

While keeping the halo and black hole mass fixed, we explore the black hole dynamics at varying dark matter mass resolution. We select an ``optimal'' gravitational softening in each case, following the prescription of \citet{power}:
\begin{equation}
    \epsilon_{\rm opt} = \frac{4 R_{200}}{\sqrt{N_{200}}},
    \label{power_eq}
\end{equation}
where $R_{200}$ is the radius enclosing 200 times the critical density of the Universe (here we take $H_0 = 100\,{\rm km\, s^{-1}Mpc^{-1}}$) and $N_{200}$ is the number of particles within this radius. The number of particles used, the black hole to dark matter mass ratio, and the optimal gravitational softening\footnote{We note that this softening is in principle 4 times larger than the actual optimal softening at $z=0$ and is actually chosen so as to minimize relaxation effects in the halo progenitor, if these simulations were cosmological.} are listed in Table~\ref{table1}. 

Our primary N-body simulations are run with the code {\small GADGET-4} \citep{gadget4}, a parallel multi-purpose cosmological N-body code that can  optionally also include gas with smoothed  particle hydrodynamics. 
We employ its fast multipole method (FMM) of order 5 for computing gravity in this work, combined with a very conservative force accuracy setting so that the influence of residual force errors is negligible in the calculations carried out here. 

\subsection{Analytical predictions}
\label{analytical_prediction_sec}

Our analytical prediction is based on applying dynamical friction with the Chandrasekhar formula (Eq.~\ref{chandrasekhar_formula}), where the distribution function is replaced
with the Hernquist distribution function, $f(r,u)$ \citep{chatterjee}. Following \citet{just_khan}, the minimum and maximum impact parameter in the Coulomb logarithm is defined as $b_{\rm min} = GM_{\rm BH}/(v_{\rm BH}^2 + 2/3 \sigma^2$), where the denominator describes the typical velocity of encounters in the vicinity of the black hole. The maximum impact parameter is defined as $b_{\rm max} = {\rho}/{|\nabla \rho|}$, where $\rho$ is the local density and $\nabla \rho$ the density gradient. For a Hernquist cusp, $b_{\rm max}$ is equal to $|r - r_0|$, the position of the black hole from the halo centre. We chose to use $b_{\rm max} = R_{\rm BH}$ in the Coulomb logarithm. We found that this choice captures accurately the behaviour of the black hole in {\small KETJU}, while setting $b_{\rm max} = \rho / |\nabla \rho | $ tends to sink the black hole substantially more slowly. This distance- and velocity dispersion-dependent choice of the Coulomb logarithm has been shown to yield a good approximation for point-like masses sinking in cuspy haloes with $\rho \propto r^{-1}$ \citep{Hashimoto_2003,  just_penarrubia, just_khan,arcasedda}. In practice, we set up a black hole particle on a circular orbit within an analytical Hernquist potential within {\small GADGET-4}, and apply an additional dynamical friction force using the above parameterisation of the Chandrasekhar formula.

 One issue with the Chandrasekhar treatment is that as the black hole approaches the centre of the halo, $b_{\rm min}$ can become greater than $b_{\rm max}$ (since $b_{\rm max}$ is defined as the black hole distance) and, as the velocity dispersion tends to zero at the centre of a Hernquist halos,  $b_{\rm min}$ becomes very large. The effect, if left untreated, can lead to black holes getting kicked  out of the halo centre when the Coulomb logarithm flips sign. We address this problem by switching off dynamical friction when this occurs. As a result, black holes do not fully sink to the centre, but settle on a circular orbit at a radius where $b_{\rm max} > b_{\rm min}$. We note that this is an issue that is also encountered by approaches like that of \citet{tremmel}, though in this case this would correspond to dynamical friction being fully resolved (the minimum impact parameter is greater than the scales over which dynamical friction is unresolved). Additionally, as simulated halos have a central density core with size $\sim \epsilon_g$, the black holes tend to settle on a circular orbit with an orbital radius similar to $\epsilon_g$.

We note that although this approach takes into account the variation in the background density of the isotropic Hernquist halo, it still applies the dynamical friction only parallel to the black hole's motion. The latter is a consequence of assuming an isotropic and homogeneous particle background in the Chandrasekhar formula, such that perpendicular contributions cancel due to symmetry. This is not the case in reality for a Hernquist halo, where a density gradient exists. Nevertheless, as we show below, this assumption seems to return a correct black hole sinking timescales when compared to high-resolution simulations. Past works have also found  that the contribution to dynamical friction due to density inhomogeneity is rather small \citep{just_penarrubia}.

\subsection{Predictions from simulations with {\small KETJU}}

It is important to note that the Chandrasekhar formula makes a number of assumptions, in particular an isotropic and homogeneous nature of the particle distribution, which leads to dynamical friction acting opposite to the direction of motion, rather than also having a perpendicular term. The background particles are also assumed to move on hyperbolic orbits, such that, for example, no particles can become bound to the black hole, which could however effectively increase its dynamical mass. The self-gravity of the wake is also ignored, as well as any fundamental changes to the density field and the distribution function that the massive perturber would induce \citep{weinberg_89, banik}. Despite these issues, the Chandrasekhar formula has been shown to provide a fairly accurate description of orbital decay in a number of set-ups \citep{velazquez, Hashimoto_2003, just_khan} and less successfully in others \citep{vdb_99,mbk}. 

In order to make sure that the Chandrasekhar description works for our set-up, we also run a high-resolution $N$-body simulation using the {\small KETJU} code \citep{ketju,Mannerkoski2023}. The initial conditions we use for this simulation are highlighted in bold in Table~\ref{table1}. The essence of the {\small GADGET-4}-based {\small KETJU} code is to add a regularized region around each massive black hole in the simulations in which the gravitational dynamics is accurately integrated without gravitational softening. Throughout this work we use the regularized region size of $2.8\,\epsilon_g$ in radius, resulting in region sizes up to a few hundreds of pc, ensuring that all interactions with the black hole in the simulations are always non-softened. The interactions between dark matter particles in the region remain softened. This allows us to capture the dynamical friction below the softening length, while also accounting for deviations from the Chandrasekhar approximation (e.g.~like the assumption of hyperbolic orbits). Moreover, since a number of previous methods used to include unresolved dynamical friction in simulations rely on the Chandrasekhar description, we aim to learn about any potential shortcomings of these methods through this test.

In {\small KETJU}, the integration around black holes (the {\small MSTAR} integrator, \citealt{Rantala2020,Mannerkoski2023}) relies on algorithmic regularization techniques \citep{Mikkola1999,Mikkola2006}, a minimum spanning tree coordinate system and the Gragg-Bulirsch-Stoer (GBS) extrapolation method \citep{Gragg1965,Bulirsch1966} to achieve an extremely high integration accuracy. The efficient two-fold parallelization of {\small MSTAR} allows for integrating up to $\sim 5000$ particles in a single regularized region. In order to properly resolve dynamical friction, {\small KETJU} requires a mass ratio of $M_{\rm BH}/M_{\rm DM} \gtrsim 100$ between the black hole and the dark matter particles \citep{ketju}. In this work, we use the mass ratio $M_{\rm BH}/M_{\rm DM} = 5000$. {\small KETJU} has been successfully applied for simulating massive black hole and black hole binary dynamics in both isolated galaxy mergers (e.g. \citealt{Rantala2018}) and cosmological zoom-in simulations \citep{Mannerkoski2022}. Note that in this work we employ only Newtonian gravity and do not include post-Newtonian terms in the equations of motion close to the black hole, something that is in principle available in {\small KETJU}, as the dark matter mass elements do not represent physical particles. For the user-given accuracy parameters of {\small KETJU}, we adopt a GBS tolerance of $\epsilon_\mathrm{GBS} = 10^{-7}$  and an integration end-time tolerance of $\epsilon_\mathrm{t} = 10^{-2}$.

\subsection{Results for simulations without added dynamical friction}

\begin{figure*}
    \centering
   
    \includegraphics[width = 2\columnwidth]{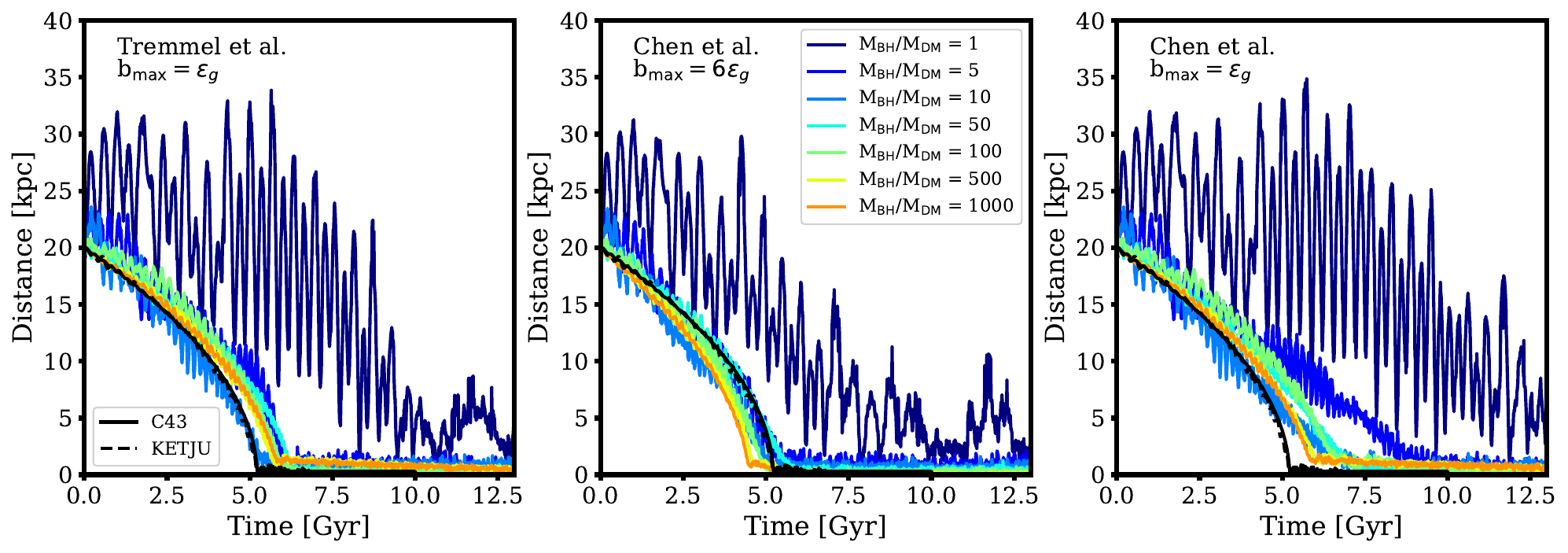} \\
   
    \caption{\textit{Left:} black hole trajectories at different halo mass resolutions after applying the \citet{tremmel} dynamical friction correction, where the relevant quantities are computed using a weighted smoothing kernel containing 100 nearest dark matter particles. The maximum unresolved impact parameter is set to $b_{\rm max,unres} = \epsilon_g$. \textit{Middle:} the result of applying the \citet{chen} dynamical friction correction, which assumes the Maxwellian approximation and $b_{\rm max,unres} = 6\, \epsilon_g$. \textit{Right:} applying the \citet{chen} correction, while setting $b_{\rm max,unres} = \epsilon_g$. }
    \label{fig2}
\end{figure*}

The left panel of Fig.~\ref{fig1} shows the results of our numerical experiment to understand the extent to which dynamical friction is resolved in simulations of different mass and spatial resolution. The solid black line shows the analytical prediction of the Chandrasekhar formula (C43) and the dashed black line is the prediction of {\small KETJU}. As expected, an increase in resolution leads to a reduction in noise, as manifested by the oscillations in black hole peri- and apocentres. For  $M_{\rm BH}/M_{\rm DM} > 10$  and $\epsilon < 2$~kpc the black hole apocentre decreases with time. This highlights that even if the mass of the black hole is greater than that of the dark matter particle, sinking is not guaranteed. 

It can also be seen that an increase in mass and spatial resolution leads to a faster sinking of the black holes. As the mass resolution and the ratio $M_{\rm BH}/M_{\rm DM}$ increase, the magnitude of the dynamical friction force becomes significantly larger than the impact of noise. A smaller gravitational softening also means a larger range of resolved impact parameters contributing to dynamical friction, thus also making the black holes sink faster.  Nevertheless, even our highest resolution run with $10^8$ particles does not converge to the analytical prediction, or the prediction of {\small KETJU}. As shown in Fig.~3 of \citet{ketju}, the gravitational softening of a {\small GADGET}-type code must be close to $\sim 1$~pc to reach convergence with the {\small KETJU} results. It is also evident that the biggest change in the sinking timescale occurs when the softening is reduced to below 1~kpc, which substantially increases the magnitude of the resolved Coulomb logarithm. Another decrease in the sinking timescale of nearly 2~Gyr occurs as the softening is reduced to $\sim 100\,{\rm pc}$, which again suggests that scales $<10 \,{\rm pc}$ must be resolved in order to obtain converged predictions. Overall, it is unlikely that the correct sinking timescale can be captured for any softening that is greater than the typical $b_{\rm min}$ of the black hole in a given halo. In our case, this implies that softening $\lessapprox 20\,{\rm pc}$ is required, which is substantially less than the ``optimal'' softening needed to suppress 2-body effects in dark matter, even in the case of 10$^8$~particles in the halo.

Evidently, gravitational softening plays a crucial role in resolving the effects of dynamical friction, and for sufficiently high mass resolution ($>10^7$ particles in the halo) its role is more important than the mass resolution itself. To confirm this, we additionally show the sinking of the black hole at each mass resolution with a fixed softening of 0.44~kpc on the right panel of Fig.~\ref{fig1}. It can be seen that the run with $M_{\rm BH}/M_{\rm DM} = 5$ is now showing some black hole sinking (although the black hole is clearly unable to sink within a Hubble time), while for mass ratios $>5$, black holes are able to sink to the centre. On the other hand, simulations with high mass resolution, which on the left panel of Fig.~\ref{fig1} had $\epsilon_g < 0.44\,{\rm kpc}$, now have delayed sinking times. We conclude that in the regime where the black hole particle mass is substantially higher than that of the dark matter the gravitational softening is more important than the mass resolution in resolving the dynamical friction force. The right panel of Fig.~\ref{fig1} also shows that for the case of $M_{\rm BH}/M_{\rm DM} = 1$, the black hole is kicked to higher apocentres when the gravitational softening is too small. This is because such a small softening is unable to effectively suppress the noise in the gravitational potential at this mass resolution, leading to the black hole getting kicked outwards (note that this also applies to the dark matter particles). 

A particularly interesting case is the simulation with 10$^5$ particles ($M_{\rm BH}/M_{\rm DM} = 10$), which, with a smaller softening, seems to have its black hole sinking on roughly the correct timescale, unlike its counterparts with higher mass resolution. The enhanced 2-body scattering in background particles due to the small softening, combined with the particulars of the initial conditions, have changed the black hole's orbit in such a way that it happens to come close to the centre of the halo early on and thus also experiences more dynamical friction early on. Despite the apparent success in representing black hole sinking in this case, we interpret this as more of a lucky coincidence and a warning against an inadequate use of gravitational softening at a given mass resolution in simulations (i.e.~noise can by chance cause black holes to sink faster, but such a behaviour is not desirable overall in our effort to obtain converged sinking timescales for black holes at different mass resolutions).

\section{Performance of existing methods}
\label{trem_chen_sec}

In the following, we will evaluate how the existing methods to correct for unresolved dynamical friction perform on our test halo, namely the method of \citet{tremmel} and \citet{chen}. We will not test the method of \citet{pfister}, which was developed for an adaptive mesh refinement code {\small RAMSES} \citep{ramses}. For both the \citet{tremmel} and the \citet{chen} methods, we compute the relevant quantities (e.g.~local density of dark matter particles, velocity of the black hole relative to the surrounding medium, velocity dispersion) at each gravity step and add the dynamical friction force from Eqs.~(\ref{equation_tremmel}) and (\ref{equation_chen}) to the force computed by the gravity solver in {\small GADGET-4}. We also note that we do not boost the dynamical mass of the black hole for our lowest-resolution case, as would be done by a factor of 3 in the approach of \citet{tremmel} and by a factor of 2 in \citet{chen}. In this section, we will only consider simulations with up to $N_{\rm DM} \leq 10^7$ particles, representative of relatively well-resolved haloes in large-volume cosmological simulations.

\subsection{The method of \citealt{tremmel}}

To estimate the local density, $\rho_{\rm tot}$, \citet{tremmel} find the 64 nearest neighbours of the black hole and use them to estimate the local density with a smoothing kernel $\rho_{\rm tot} = \sum_i^N m_i W(|r - r_{\rm BH}|, h)$. The mass of particles within the kernel that are moving slower than the black hole is counted and their density, $\rho(v<v_{\rm BH})$, is computed as $\rho_{\rm tot} M(v<v_{\rm BH})$/$M_{\rm tot}$. The velocities are taken as those relative to the centre of mass velocity of the neighbouring particles. In our application of the method, we instead compute the relevant quantities over 100 nearest particles to reduce the particle noise, and we take $W$ as the standard {\small GADGET-4} smoothing kernel, with $h$ being the distance to the furthest neighbouring particle (see Eq.~108 of \citealt{arepo}). We compute the centre-of-mass velocity and velocity dispersion as smoothed quantities, however, we have also verified that this choice does not noticeably affect the results of the simulations. As in \citet{tremmel}, $b_{\rm max}$ in the Coulomb logarithm is taken to be the gravitational softening of each simulation.

The results of this experiment can be seen in the top left panel of Fig.~\ref{fig2}, where the black holes, on average, sink slower than predicted by {\small KETJU} by almost 1~Gyr. One exception to this is the black hole with $M_{\rm BH}/M_{\rm DM} = 10$, which, due to noise, approaches within 10~kpc of the centre of the halo already at $\sim3$~Gyr, leading to enhanced dynamical friction. Interestingly, the $M_{\rm BH}/M_{\rm DM} = 1$ black hole, which would show no noticeable dynamical friction without a correction, is also able to sink. We interpret this as a result of $b_{\rm max}$ being extremely large in this simulation (14~kpc), such that the total dynamical friction force is able to overcome the stochastic heating. Otherwise, the sinking time of the black holes is approximately converged between different resolution levels.

\begin{figure}
    \centering
    \includegraphics[width = \columnwidth]{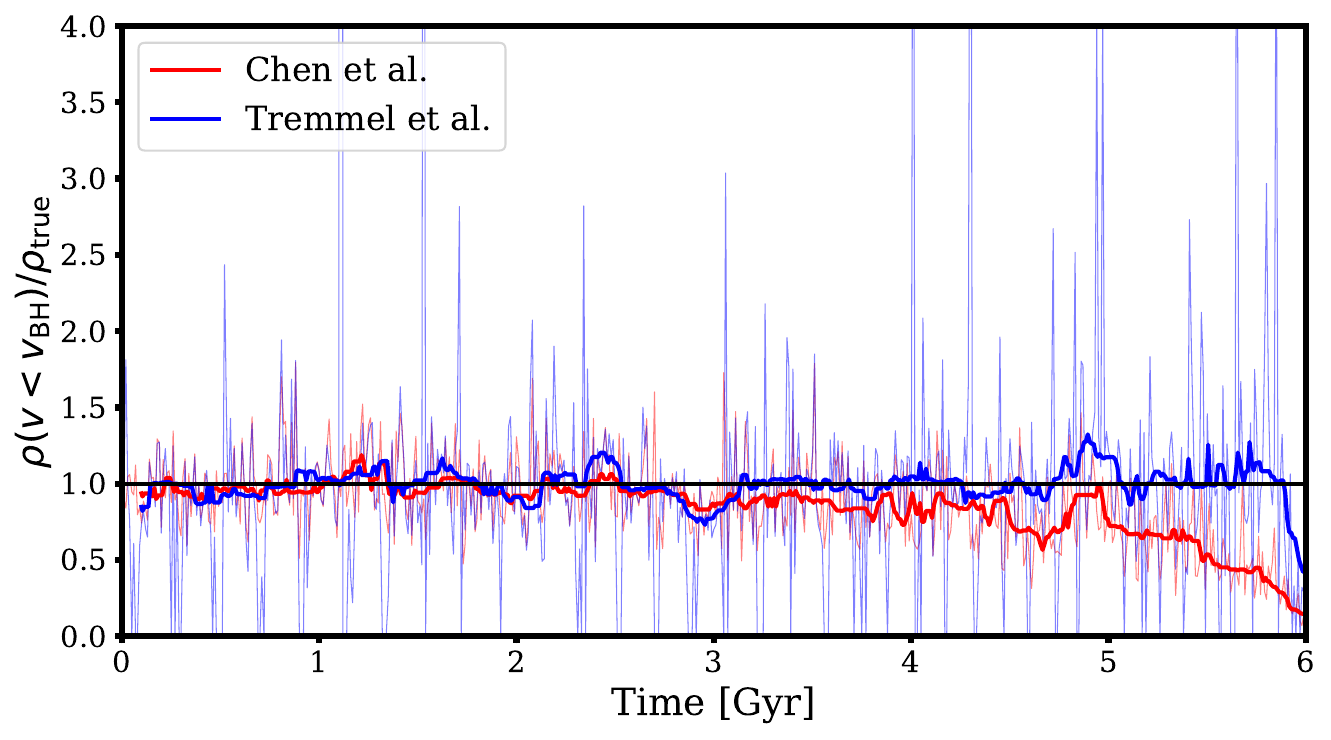}
    \caption{The estimated density of particles moving slower than the black hole, $\rho(v<v_{\rm BH})$, compared to the density derived from the Hernquist distribution function, $\rho_{\rm true}$. The black line shows the prediction, the thin red line corresponds to the use of the Maxwellian approximation, as in \citet{chen}, and the thin blue line is the kernel-smoothed estimate of \citet{tremmel}. The thick lines are moving averages over 0.2~Gyr. Both densities are estimated using a kernel size of 100 particles in a simulation with $10^6$ dark matter particles. }
    \label{fig3}
\end{figure}

\subsection{The method of \citealt{chen}}

The approach of \citet{chen} builds on that of \citet{tremmel}, where now the kernel size over which the distribution function is estimated is based on the SPH kernel, defined by neighbouring gas particles\footnote{We also note that, while gas may be a good tracer of the local dark matter density and will act as an adaptive definition of the number of particles within the smoothing kernel, the approach of \citet{chen} will likely fail in cases where the gas has been fully consumed by star formation or blown out by feedback, such as in low-mass galaxies, leading to artificially low estimates of the local density in such situations. } in cosmological simulations, rather than a fixed number of collisionless particles, which helps to reduce the noise in the density estimate. The maximum unresolved impact parameter, $b_{\rm max, unres}$, is now defined as some factor $\eta$ times the gravitational softening, where the optimal value according to their experiments is $\eta = 6$. It is certainly true that the exact value of the gravitational softening is not necessarily equal to the full extent of the unresolved dynamical friction scale. In many implementations of the gravitational softening, the forces are softened up to a distance of $\sim 2.8 \times \epsilon_g$, beyond which they are fully Newtonian \citep{arepo}. As such, we expect that increasing the value of $b_{\rm max}$ above $\epsilon_g$ will likely lead to a compensation of more of the unresolved dynamical friction and, in principle, black hole sinking rates which are more accurate.

The result of applying the \citet{chen} formula can be seen in the middle panel of Fig.~\ref{fig2}, where the black holes now seem to sink, on average, faster than the prediction of {\small KETJU} by up to 1~Gyr, and there appears to be a slight resolution-dependent gradient in the sinking time. This may suggest that $b_{\rm max} = 6\, \epsilon_g $ in the Coulomb logarithm is too large of a dynamical friction correction, particularly in cases where dynamical friction is to a large part resolved. 

To understand the effect of the enhanced $b_{\rm max}$ in the Coulomb logarithm, we now repeat our test with the \citet{chen} formula, but setting $b_{\rm max,unres} = \epsilon_g$. This is shown in the right panel of Fig.~\ref{fig2}. It can be seen that at the highest resolution of $M_{\rm BH}/M_{\rm DM} = 1000$, the sinking of the black hole is very similar to \citet{tremmel}, while lower mass-ratio black holes sink too slowly. Thus, the maximum unresolved impact parameter $b_{\rm max, unres}$ must be artificially increased by a factor of 6 to strengthen the effect of dynamical friction in these cases. Together with the shallower descent of the black holes that we see on the right panel of Fig.~\ref{fig2}, this suggests that the Maxwellian approximation delays the sinking of the black holes, particularly when the resolution is low. Previous works have pointed out this inapplicability of the Maxwellian approximation to the centres of isotropic Hernquist haloes \citep{rvine}. 

In Fig.~\ref{fig3} we show the accuracy of the $\rho(v<v_{\rm BH})$ estimates using the Maxwellian approximation (red) and using the kernel-weighted density of particles moving slower than the black hole (blue). Both are applied at each time step to the black hole with $M_{\rm BH}/M_{\rm DM} = 100$. We can see that the Maxwellian approximation is substantially more smooth than the approach of \citet{tremmel} and, on average, slightly underestimates the density of slow-moving particles. This underestimation is significant when the black hole descends to below 10~kpc (> 3~Gyr). 

We additionally note that setting $b_{\rm max,unres} = 6\,\epsilon_g$ appears to be more effective in keeping the black holes in the centre of the halo (middle panel of Fig.~\ref{fig2}, compared to the left and right panels). This is because the condition $b_{\rm min} > b_{\rm max,unres}$ sets the dynamical friction correction to zero. Thus, the higher $b_{\rm max, unres}$  is, the more do the black holes move towards the local dark matter density centre. Another advantage of using the Maxwellian approximation is that it ensures $\rho(v<v_{\rm BH}) > 0$ at all times, such that there is always a dynamical friction force, provided $b_{\rm min} < b_{\rm max, unres}$. We note however, that this may not be beneficial in shallow density profiles or dark matter cores, where no background particles are moving slower than the black hole, resulting in delayed sinking of the black hole or core stalling \citep{read, petts}.

We conclude that the method of \citet{tremmel} tends to sink the black holes too slowly, suggesting that the condition $b_{\rm max} = \epsilon_g$ is insufficient to account for unresolved dynamical friction. The method of \citet{chen} tends to also underestimate the dynamical friction, due to the use of the Maxwellian approximation, particularly in the inner regions of an isotropic Hernquist halo and thus requires an enhanced value of $b_{\rm max,unres}$ to compensate for this underestimation.  We confirm that $b_{\rm max,unres} = 6\,\epsilon_g$ is an effective correction at low resolution, but at high resolution it leads to a too fast sinking of black holes. 

\section{A calibrated dynamical friction model}
\label{my_model_sec}

\begin{figure}
    \centering
    \includegraphics[width=\columnwidth]{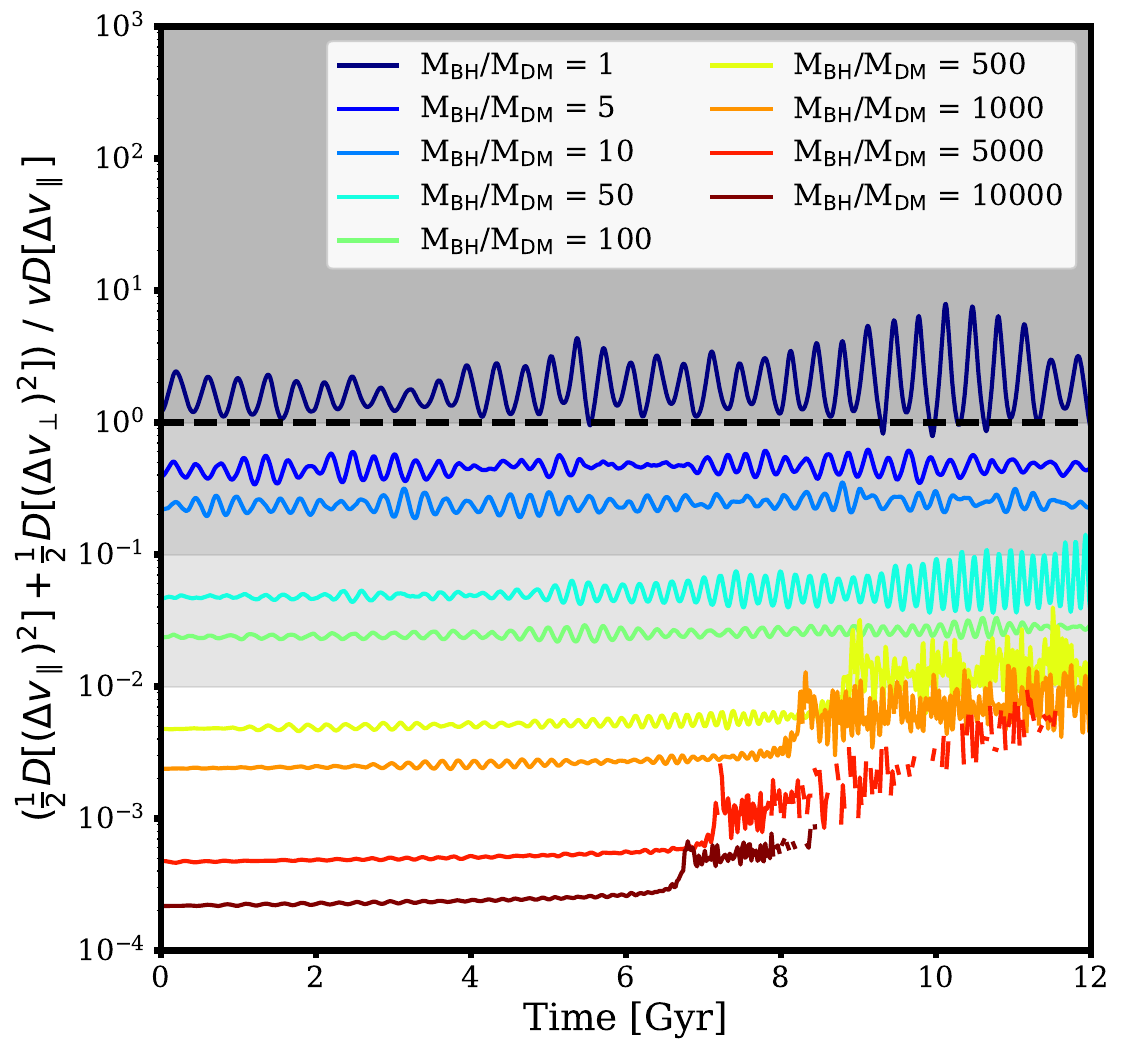}
    \caption{The contribution of the second-order diffusion coefficients to the change in the kinetic energy of the black hole,$\frac{1}{2}D[(\Delta v_{\parallel})^2] + \frac{1}{2}D[(\Delta v_{\perp})^2]$, compared to the contribution of dynamical friction, $v D[\Delta v_{\parallel}]$. The ratio of energy contributions is computed as a function of time for the black hole orbits shown in the left panel of Fig.~\ref{fig1}, but assuming unsoftened interactions. The terms are computed through evaluation of Eqns.~(\ref{cooling_eq}), (\ref{heating_eq}) and (\ref{heating_eq2}) for a Hernquist halo. The black dashed line shows where the contributions of the two terms are equal. The progressively darker shaded regions indicate where the contribution of stochasticity is above the 1 and 10 and 100~percent level of the dynamical friction contribution. Orbits of black holes with mass ratios as high as $M_{\rm BH}/M_{\rm DM}=100$ can be substantially altered by stochastic effects (see also Fig.~\ref{fig11}).}
    \label{fig4}
\end{figure}

As we have seen in the previous section, dynamical friction estimators like those of \citet{tremmel} and \citet{chen} are able to model the long-term orbital evolution of individual black holes, rather than sinking them immediately as in repositioning methods, and, with the right choice of parameters, these methods can sink black holes to the centres of their galaxies on a correct timescale. We will therefore adopt a similar form for our subgrid dynamical friction model:
\begin{equation}
{\bf a}_{\rm df,M_{\rm BH}} = - \frac{4 \pi G^2 \rho(v < v_{\rm BH}) }{v^3}  M_{\rm BH} \ln\frac{\eta \epsilon_g}{b_{\rm min}} {\bf v},
\label{mymodel}
\end{equation}
 where $\eta$ is a factor by which the softening is multiplied to obtain the maximum unresolved impact parameter. This is the value that we will attempt to calibrate through comparison with {\small KETJU}.

Another choice we must make is the definition of the minimum impact parameter, $b_{\rm min}$. Given the success of the Chandrasekhar analytical prediction in Section~\ref{analytical_prediction_sec} compared to that of {\small KETJU}, we choose to define the minimum impact parameter as:
\begin{equation}
{b_{\rm min}} = \frac{GM_{\rm BH}}{(v^2_{\rm BH} + \frac{2}{3}\sigma^2_{\rm 3D})}, 
\label{equation8}
\end{equation}

which additionally sets the ``true'' maximum impact parameter $b_{\rm max} = R_{\rm BH}$.

\subsection{Dynamical friction vs. stochasticity in collisional systems}
\label{diff_coeff}
Since we are attempting to reproduce the sinking of a black hole as it would occur in a smooth dark matter potential, where the masses of background particles tend to zero, we must be wary of how discreteness can affect black hole dynamics. Dynamical friction is one of the effects arising from gravitational encounters, which causes a black hole to lose its kinetic energy; however, scattering by massive background particles can also increase the kinetic energy of a black hole.

In a collisionless system, the phase-space density, $f$, around a test particle does not change with time. However, if gravitational collisions are present, the phase space density changes at a rate described by the encounter operator $\Gamma[f]$. One way to obtain the Chandrasekhar formula is via the diffusion coefficients of the Fokker-Planck approximation to the encounter operator \citep{spitzer,binneytremaine}. Namely, the first diffusion coefficient
\begin{equation}
 D[\Delta v_{\parallel}] = -\frac{16 \pi^2 G^2 (M_{\rm BH} + m_{*} )\ln \Lambda}{v^2} \int_0^v {\rm d}v_{*} v^2_{*}f({\bf x}, v_*)  
 \label{cooling_eq}
\end{equation} 
 is just the dynamical friction formula of Eq.~(\ref{chandrasekhar_formula}). The method of, e.g., \citealt{tremmel} neglects the additional ``cooling'' term arising from the mass $m_*$  of the background particles and thus assumes $m_*\ll M_{\rm{BH}}$. The method of \citet{damiano} includes the $m_*$ term in their correction for unresolved dynamical friction; however, when the black hole mass is lower than the dark matter particle mass, the resultant dynamical friction force is dominated by the $m_*$ term, which would otherwise be negligible in a well-resolved system\footnote{This explains why the correction of \citet{damiano} is relatively efficient in sinking low-mass black holes i.e. they assume that the background star and dark matter particle masses in simulations are the ``true'' masses. }. 
 
 For sufficiently large mass of the background particles compared to the black hole, the two second-order diffusion coefficients become important. The first second-order diffusion coefficient,
\begin{multline}
    D[(\Delta v_{\parallel})^2] = \frac{32 \pi^2 G^2 m_{*} \ln \Lambda}{3} \times \\ \left[ \int_0^v {\rm d}v_{\rm *} \frac{v^4_{*}}{v^3}f({\bf x}, v_*)~+  \int_{v}^{\infty} {\rm d}v_{*} v_{*} f({\bf x}, v_*) \right] ,
    \label{heating_eq}
\end{multline} 
describes the stochastic contribution to the change of velocity squared per unit time in the direction of motion of the black hole. The second second-order coefficient,
\begin{multline}
    D[(\Delta v_{\perp})^2] = \frac{32 \pi^2 G^2 m_{*} \ln \Lambda}{3} \times \\ \left[ \int_0^v {\rm d}v_{\rm *} \left(  \frac{3v_*^2}{v}-\frac{v^4_{*}}{v^3} \right)f({\bf x}, v_*)~+  2\int_{v}^{\infty} {\rm d}v_{*} v_{*} f({\bf x}, v_*) \right] ,
    \label{heating_eq2}
\end{multline} 
describes the heating in the direction perpendicular to the black hole motion\footnote{This is distinct from the perpendicular dynamical friction term of \citet{ma_hopkins}, which arises from radial inhomogeneity of the dark matter halo. We do not take this term into account in this work.}.

 We note that the order of magnitude of the second-order coefficients is determined primarily by the mass of the background particles, while the first-order coefficient (the strength of dynamical friction) depends on the black hole mass. Thus, the overall effect of stochastic heating on the sinking timescale of the black hole depends sensitively on the ratio $M_{\rm BH}/M_{\rm DM}$. The coefficient in Eq.~(\ref{heating_eq}) acts opposite to the direction of the dynamical friction force and accounts for the velocity dispersion of the slow-moving background particles and the typical velocity of the fast-moving particles.
 
 In Fig.~\ref{fig4} we show the contribution of the second-order diffusion coefficients to the energy change of the black hole compared to the contribution of the first-order (dynamical friction) coefficient. Overall, the stochastic terms become important for low $M_{\rm BH}/M_{\rm DM}$ and, particularly, in regions of the halo where the velocity dispersion of background particles is larger than the black hole velocity. We note that even at $M_{\rm BH}/M_{\rm DM}=100$, stochasticity can increase the kinetic energy of the black hole by a few per cent (at each timestep). Over time, these changes can add up and substantially alter the orbit of the black hole. It is therefore important that we are aware of the effects of the noise in the context of our model.

Therefore, in order to recover the dynamical friction force on a black hole in a smooth potential ($M_{\rm DM} \xrightarrow[]{} 0$) from the simulations, the full correction should take the form:
\begin{equation}
{\bf a}_{\rm corr} =  {\bf a}_{\rm df, M_{BH}} - {\bf a}_{\rm df, m_*} - {\bf a}_{\rm noise, \parallel} - {\bf a}_{\rm noise, \perp},
\end{equation}
with ${\bf a}_{\rm noise, \parallel}$ described by:
\begin{equation}
{\bf a}_{\rm noise, \parallel} = \frac{4 \pi G^2 \rho(v < v_{\rm BH}) }{v^3} m_*\frac{\sigma_{\rm 3D}^2}{3v^2}\ln\frac{R}{\eta \epsilon_g} {\bf v} ,
\label{noise_corr}
\end{equation}
where we have adopted the Maxwellian approximation in Eq.~(\ref{heating_eq}) \citep{binneytremaine,merritt_book}.

Through explicit evaluation of Eqns.~(\ref{cooling_eq}), (\ref{heating_eq}) and (\ref{heating_eq2}), we have determined that for an isotropic Hernquist halo the second-order diffusion term acting parallel to the direction of motion of the black hole, ${\bf a}_{\rm noise, \parallel}$, approximately cancels out the additional cooling term, ${\bf a}_{\rm df, m_*}$, and only dominates near the centre of the halo. The second-order perpendicular term, ${\bf a}_{\rm noise, \perp}$, is approximately equal in magnitude to the parallel term, suggesting that Eq.~(\ref{noise_corr}) is a good approximation to the magnitude of the noise terms (see Appendix A). However, we also found that the Maxwellian approximation typically overestimates the stochastic term in a Hernquist halo by a factor of $\sim 3$, while the numerical evaluation of local $\sigma_{\rm 3D}/v$ is extremely noisy and can lead to order-of-magnitude changes in the black hole acceleration. Moreover, it is unclear how one would account for the second-order terms perpendicular to the black hole motion, as we cannot at any given point determine the net direction of the stochastic force.

For these reasons, we do not include the correction in Eq.~(\ref{noise_corr}) in this work. Instead, in the following we will derive an optimal value of $\eta$ that in effect marginalizes over these additional effects. For $M_{\rm BH}/M_{\rm DM} > 100$, the stochasticity contribution is at sub-per-cent level and will thus not influence our results (i.e. the optimal $b_{\rm max,unres}$ should only depend on the \textit{spatial} resolution). Nevertheless, as we will discuss in Section~\ref{sec_low_mass}, the stochastic effects can be substantial for $M_{\rm BH}/M_{\rm DM} \leq 100$. We  will therefore adopt an order-of-magnitude approximation for the {\it overall} effect of the second-order diffusion coefficients.

\subsection{Calibrating the maximum unresolved impact parameter}
\label{themethod}
An obvious approach to calibrating $\eta$ would be to directly probe a range of values of this parameter by running the set of simulations from Table~\ref{table1} with the corresponding dynamical friction correction. However, in this work we take a different approach that allows us to probe a wide range of $\eta$ at low computational cost. Our procedure is as follows:

\begin{enumerate}
\item
For each simulation with no dynamical friction prescription (left panel of Fig.~\ref{fig1}), we take the position, velocity and acceleration of the black hole in the direction of motion, $a_{\rm GADGET4}$, at each timestep (excluding the times after black hole crosses within 2~kpc of the halo centre, where the density profile is suppressed compared to the Hernquist formula and where, at higher resolution, we would expect the local density centre of the halo to move in response to the motion of the black hole).

\item For a black hole inside an analytic Hernquist potential, with the same position and velocity vectors as the simulations in step~(i), we compute the expected acceleration in the direction of motion of the black hole due to dynamical friction together with the effect of the external potential, $a_{\rm analytical}$, following Eq.~(\ref{chandrasekhar_formula}).  

\item In a dark-matter-only simulation of the same halo and with the same gravitational softening, we measure the expected noise in the acceleration at the position of the black hole, $\sigma_{\rm GADGET4}$, over multiple snapshots to account for noise due to the initial conditions. The noise arises here due to the discreteness of the $N$-body potential and we assume it to be isotropic at any given location.

\item We sample values of $b_{\rm max,unres}$ in the range 0-100~kpc, and apply dynamical friction via Eq.~(\ref{mymodel}), $a_{\rm Eq.13}$. We assume that the density of particles moving slower than the black hole as well as the~3D~velocity dispersion are known exactly. Using the affine invariant Markov Chain Monte Carlo ensemble sampler {\small EMCEE} \citep{emcee}, for each $b_{\rm max, unres}$ sample we compute the log-likelihood:
\begin{equation}
\ln L  = -\frac{1}{2}\sum_{t=0}^{t=\mathrm{end}} \frac{({a_{\mathrm{GADGET4}} + a_{\mathrm{Eq.13}} - a_{\mathrm{analytical}}})^{2}}{\sigma_{\mathrm{GADGET4}}^2}.
\end{equation}

\item After a Monte Carlo Markov Chain (MCMC) burn-in length of 2500 steps with 250 chains ({\small EMCEE} walkers), we use the next 2500 steps to estimate the distribution of optimal $b_{\rm max, unres}$ choices.
\end{enumerate}

We note that, as an alternative method, we could instead determine the minimum \textit{resolved} impact parameter $b_{\rm min, res}$ in simulations, given $b_{\rm max} = R$. This method is equivalent to our approach and we checked that this produces identical results. We additionally emphasise that the crucial aspect of this procedure is having a well-defined analytical $b_{\rm min}$ and $b_{\rm max}$ in the Coulomb logarithm. The former is well defined for a point particle as the impact parameter for a 90-degree deflection, while the latter is constrained through comparison with {\small KETJU}.

\begin{figure}
    \centering
    \includegraphics[width=\columnwidth]{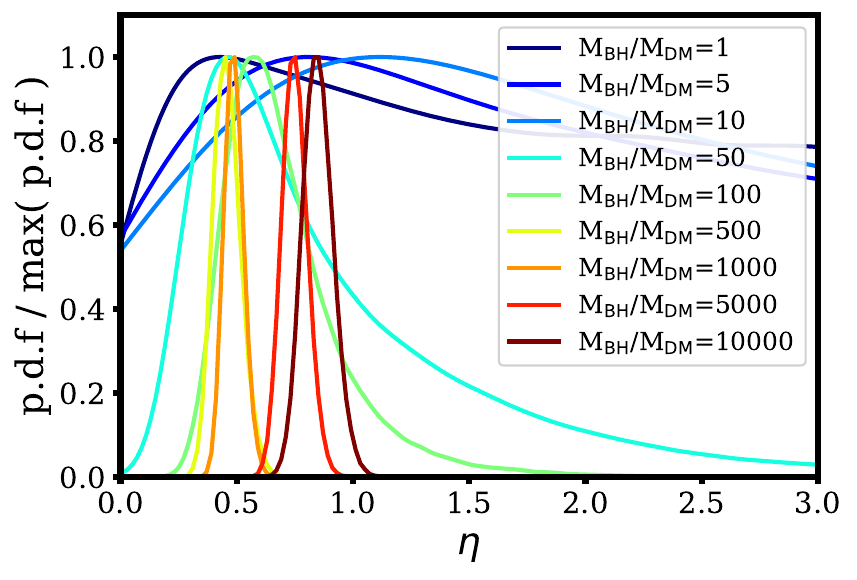}
    \caption{{\small EMCEE}-generated samples of the distribution of optimal values of $\eta = b_{\rm max}/\epsilon_g$ for the set of simulations in Table~\ref{table1}. The results are shown for the range $\eta = 0 - 3$ for clarity. No convergence is reached for $M_{\rm BH}/M_{\rm DM} \leq 10$, though the peak of the distribution lies below $\eta\approx1$. The optimal values of $\eta$ differ as the mass resolution and softening changes. Higher-resolution simulations prefer $\eta = 0.5-0.8$.}
    \label{fig5}
\end{figure}

The outcome of this exercise can be seen in Fig.~\ref{fig5}, where we show the resulting distributions of $\eta = b_{\rm max, unres}/\epsilon_g$. It is clear from Fig.~\ref{fig5} that for the mass ratio $M_{\rm BH}/M_{\rm DM} \leq 10$ our method has failed to converge on an optimal $\eta$, although already at $M_{\rm BH}/M_{\rm DM} = 5$ we can see that this value needs to be above zero. This result is not surprising, as at the lowest resolution, $M_{\rm BH}/M_{\rm DM} = 1$, we do not expect to see any effects of dynamical friction. In this case, the orbit of the black hole is quite noisy, and in some instances the dynamical friction correction would need to be either small or large, and positive or negative, to obtain the acceleration that corresponds to the analytical prediction. This is still partially the case until $M_{\rm BH}/M_{\rm DM} > 10$. At $M_{\rm BH}/M_{\rm DM} \geq 100$, however, we are already able to effectively narrow down the range of optimal values of $\eta$, which appears to be within $0.5-0.7$.

We also note that in our approach of examining the simulations from the left panel of Fig.~\ref{fig1}, some of the black holes (particularly at the three lowest resolution levels) never approach the centre of the halo. This could suggest that the lack of a constraint on $\eta$ at this resolution arises due to this missing data; however, we have explored the radial dependence of our constraint (i.e.~repeating the same procedure as outlined above but for radial bins of 2~kpc at all resolution levels) and found consistent best values of $\eta$ down to the inner $\sim2\epsilon_g$, where the density profiles of $N$-body haloes would naturally deviate from analytical predictions.  

\subsection{Probing different values of the softening}
\label{grid_section}

\begin{figure*}
    \centering
    \includegraphics[width = 2\columnwidth]{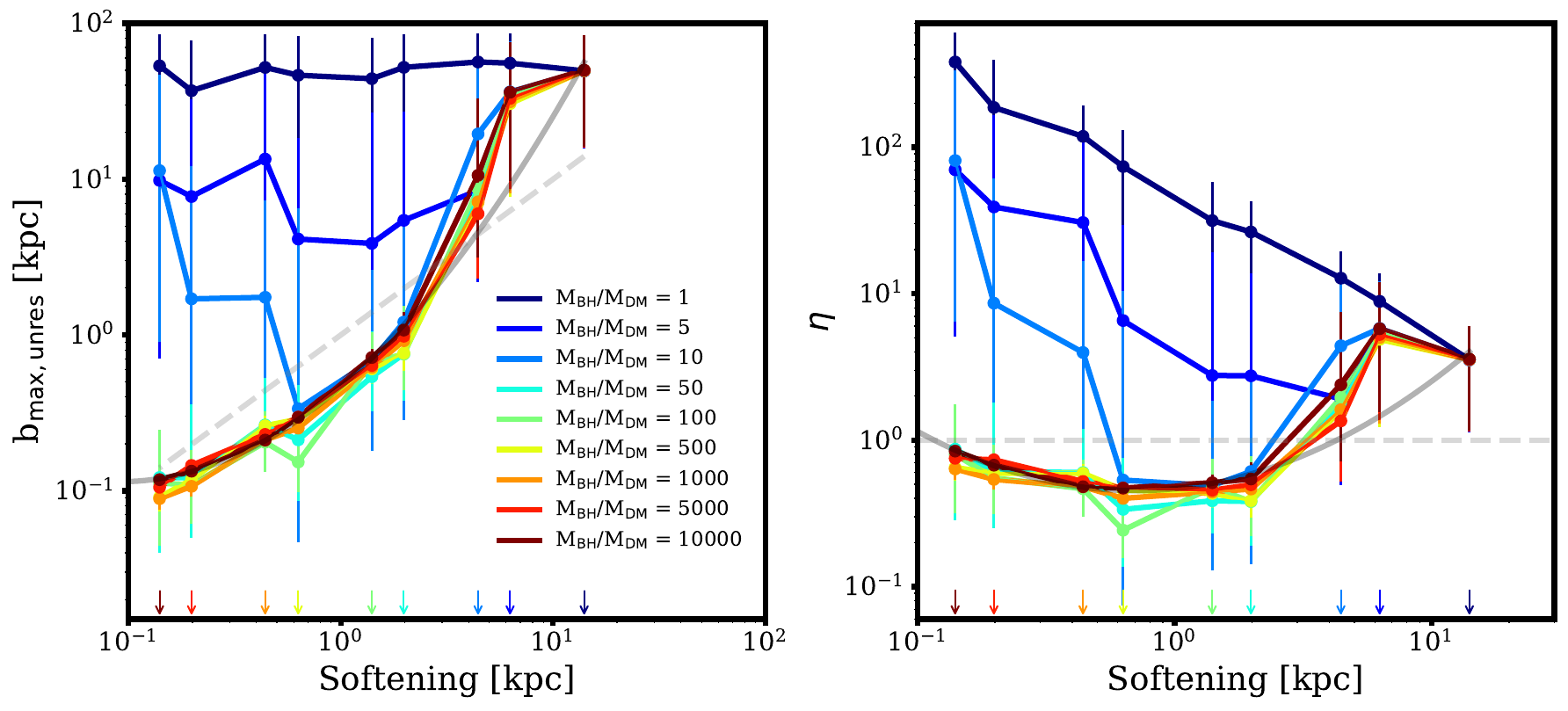}
    \caption{\textit{Left:} the maximum unresolved impact parameter, $b_{\rm max, unres}$, as a function of gravitational softening, $\epsilon_g$, obtained through the method described in Section~\ref{themethod}. Different colours correspond to the mass resolution of a $10^{13}\, {\rm M}_\odot$ halo hosting a $10^{9}\,{\rm M}_\odot$ black hole and the respective black hole to dark matter particle mass ratio. For each of the 9 levels of mass resolution, simulations of the $10^{9}\, {\rm M}_{\odot}$ black hole starting off at 20~kpc on a circular orbit were run for 9 different values of the softening. The circles indicate the median of the posterior distribution of the maximum unresolved impact parameter obtained with {\small EMCEE} for each simulation, and error bars show the 16$^{\rm th}$ and 84$^{\rm th}$ percentiles. The solid grey curve shows the least-squares fit of a quadratic relation to all data points with M$_{\rm BH}/M_{\rm DM} > 1$. The dashed grey line shows $b_{\rm max, unres} = \epsilon_g$ \citep{tremmel}. The arrows indicate the optimal softening at each mass resolution, according to \citet{power}. \textit{Right:} same as the left panel, but for the constrained optimal $\eta=b_{\rm max, unres}/\epsilon_g$.}
    \label{fig6}
\end{figure*}

Our experiment in Fig.~\ref{fig5} was limited to the range of ``optimal'' softening choices according to \citet{power}, and it is not yet clear whether the required optimal $\eta$ would be the same if the softening was chosen differently. We therefore repeat the procedure outlined in the previous section, now varying the softening for each background particle mass resolution. The softening values we use at each resolution level are the same as those in Table~\ref{table1}.

The result of this is given in Fig.~\ref{fig6}, where on the left we show the maximum unresolved impact parameter as a function of the softening, and on the right we show the optimal $\eta$ obtained through our method in each case. Firstly, we see that for all gravitational softening values there is no constraint on $b_{\rm max, unres}$ for the lowest resolution run, with $M_{\rm BH} = M_{\rm DM}$. At high resolution ($M_{\rm BH}/M_{\rm DM} \geq 50$ and $\epsilon_g \leq 2$) there appears to be a clear correlation between the softening length and unresolved $b_{\rm max}$, as well as a convergence around the optimal value of $\eta$.  Above $\epsilon_g = 2$~kpc, the relation appears to steepen, however within the error bars it is still consistent with the linear relation we see for higher-resolution simulations. The steepening of the relation is likely reflective of the unresolved central densities of the simulated halos, thus requiring stronger dynamical friction to sink the black holes consistently with the analytical prediction.

The results for $M_{\rm BH}/M_{\rm DM} = 5$ and $M_{\rm BH}/M_{\rm DM} = 10$ have a particularly interesting shape, where at larger values of softening the relation is followed, while at values of the softening that are too low for this resolution, $\epsilon_g < 2$~kpc, the maximum unresolved $b_{\rm max}$ appears to increase. We interpret this as the result of enhanced two-body interactions that introduce additional noise in the black hole orbit. Nevertheless, within the errors, even these results are consistent with the general relation seen at higher resolution. We thus see no evidence of a relation between the mass resolution and the maximum unresolved values of $b_{\rm max}$, which must be set only by the gravitational softening.

\begin{figure}
    \centering
    \includegraphics[width = \columnwidth]{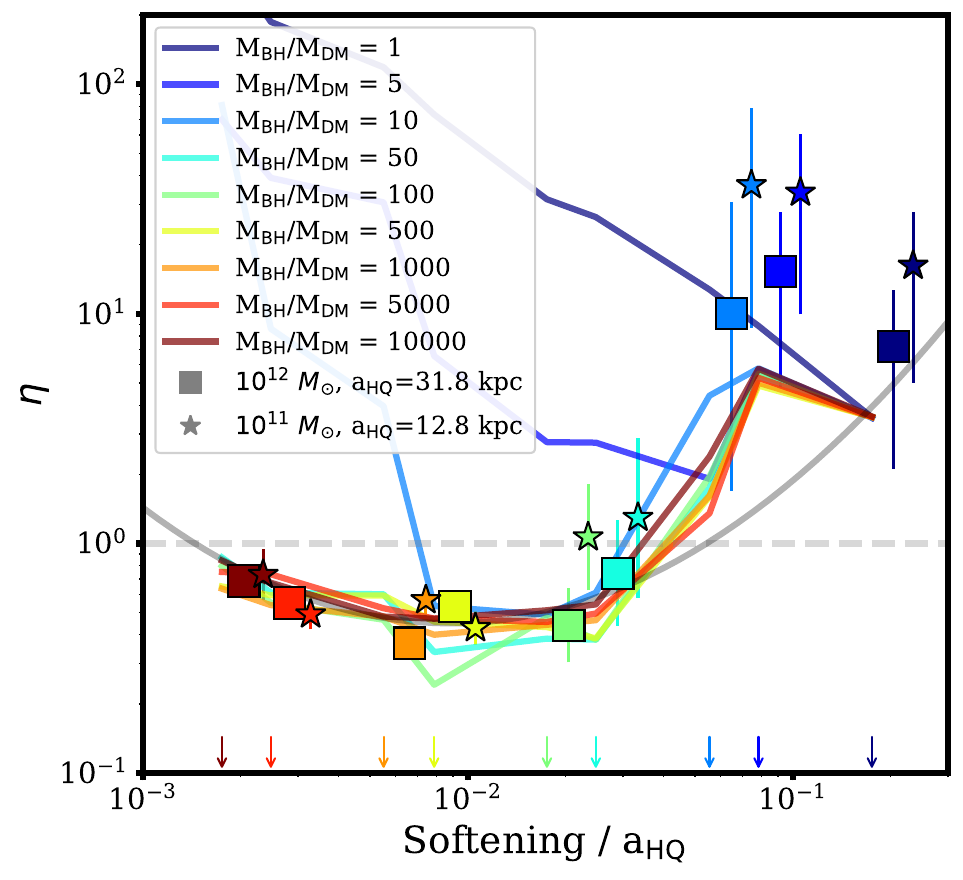}
    \caption{Similar to Fig.~\ref{fig6}, but showing our results for black holes and haloes of varying mass. The lines show the optimal $\eta$ at each mass resolution for our fiducial $10^{13}\,{\rm M}_{\odot}$ halo as a function of softening, normalized by the scale length of the Hernquist halo ($a_{\rm HQ} = 80$~kpc). The colours denote the mass resolution and arrows give the optimal softening at a given mass resolution for the $10^{13}\,{\rm M}_{\odot}$ halo. Square and star symbols show derived values of $\eta$ at optimal softening for a $10^{12}\, {\rm M}_{\rm \odot}$ halo ($10^{8}\, {\rm M}_{\rm \odot}$ black hole starting at 10~kpc) and a $10^{11}\,{\rm M}_{\rm \odot}$ halo ($10^{7}\,{\rm M}_{\rm \odot}$ black hole starting at 5~kpc), respectively. }
    \label{fig7}
\end{figure}

To obtain the optimal value of $\eta$, we fit a quadratic relation to the left panel of Fig.~\ref{fig7}. This is shown with a solid grey line and corresponds to the best fit value
\begin{equation}
\eta_{\rm optimal} = \rm min[ 0.46~({\epsilon_g}/{\rm kpc})^{0.17}~10^{~0.11 \ln^2(\epsilon_g/{\rm kpc})}~,~1.0 ], 
\label{optimal_eta_13}
\end{equation}
where we fixed the maximum value of $\eta$ to 1. The relation $b_{\rm max} = \epsilon_g$ is shown with the grey dashed line. We note that the definition of $b_{\rm min}$ of \citet{tremmel} and \citet{chen} is different to ours, so our relation would not hold in that case; however, with $b_{\rm min} = GM/v^2_{\rm BH}$, and through an approximation $v^2_{\rm BH}\approx \frac{1}{3}\sigma_{\rm 3D}^2$, we can deduce that an optimal value is $\eta \approx 2$ for \citet{tremmel} and \citet{chen}. We show the optimal value of $\eta$ with this definition of $b_{\rm min}$ in Appendix B.

Finally, we wish to verify that our relation is not specific to the particular black hole mass and its corresponding orbit within our mock halo. To do this, we now repeat our test, while varying the halo and black hole properties. The result of this can be seen in Fig.~\ref{fig7}, where we show constraints for the optimal vale of $\eta$ for a $10^8\,{\rm M}_{\odot}$ black hole in a $10^{12}\,{\rm M}_{\odot}$ halo, and a $10^7\,{\rm M}_{\odot}$ black hole in a $10^{11}\,{\rm M}_{\odot}$ halo. It can be seen that the obtained values agree well with our previously derived relation when the softening is normalized by the Hernquist scale radius $a_{\rm HQ}$, suggesting that the variation in optimal $\eta$ is dependent on the rate of encounters experienced by the black hole in the simulation. 

The generalized relation for any given Hernquist-like halo may therefore be expressed as:
\begin{equation}
\eta_{\rm optimal} = \rm min[ 103.7~({\epsilon_g/a_{\rm HQ}})^{1.3}~10^{~0.11 \ln^2(\epsilon_g/a_{\rm HQ})}~,~1.0 ].
\label{optimal_eta_scaled}
\end{equation}
We note that while $a_{\rm HQ}$ can be estimated with an on-the-fly halo finding algorithm, in practice, many cosmological simulations will not reach the softening of $~10^{-3}a_{\rm HQ}$ and will thus fall on the `flat' part of our relations with $\eta \approx 0.45 \epsilon_g$. 

\subsection{The meaning of the optimal $\eta$}

It may at first seem surprising that our derived optimal values of $\eta$ are smaller than 2.8$\epsilon_g$, given that this is the radius where the forces become fully Newtonian in {\small GADGET-4} (i.e. particle scattering follows the full Newtonian case). To gain an intuitive understanding of this apparent discrepancy, we consider 2-body gravitational scattering. The use of the gravitational softening introduces a maximum scattering angle in a given interaction. The impact parameters that correspond to stronger particle deflections are indeed unresolved (for the spline softening we infer that the maximum deflection occurs at $b \simeq 1.4\,\epsilon_g$, see \citealt{white1976, ludlow_impact}). However, this also implies that there is now an excess of particles with weaker deflections, effectively increasing the rate of encounters with impact parameters $b > 1.4\,\epsilon_g$ (though the exact dynamical friction effect will depend on the momentum exchange resulting from the scattering). Therefore, the correction to dynamical friction that one needs to apply should not only compensate for the unresolved impact parameters but also take away the excess encounters with weaker deflections, which is what our relation achieves.

A seemingly puzzling feature of our recovered relation is that, while the maximum unresolved impact parameter $b_{\rm max, unres}$ does decrease with the softening, it appears to decrease slower towards smaller values of the softening, resulting in a $\eta(\epsilon_g)$ relation that has a parabolic shape. However, we can also obtain a similar parabolic relation using equation~6 of \citet{white1976}, which computes the minimum impact parameter for an interaction of an extended object (e.g. a globular cluster) with a field star, where for the mass distribution of the subject we assume the distribution implied by the spline softening in {\small GADGET-4}. We note however, that the treatment of \citet{white1976} assumes undeflected trajectories of field stars when computing their velocity change due to the gravitational encounter with an extended object and thus only sets an upper limit on their minimum impact parameter ($b_{\rm min} \leq 1.4\, \epsilon_g$).

\subsection{Optimizing the kernel size}

\begin{figure*}
    \centering
    \includegraphics[width=2\columnwidth]{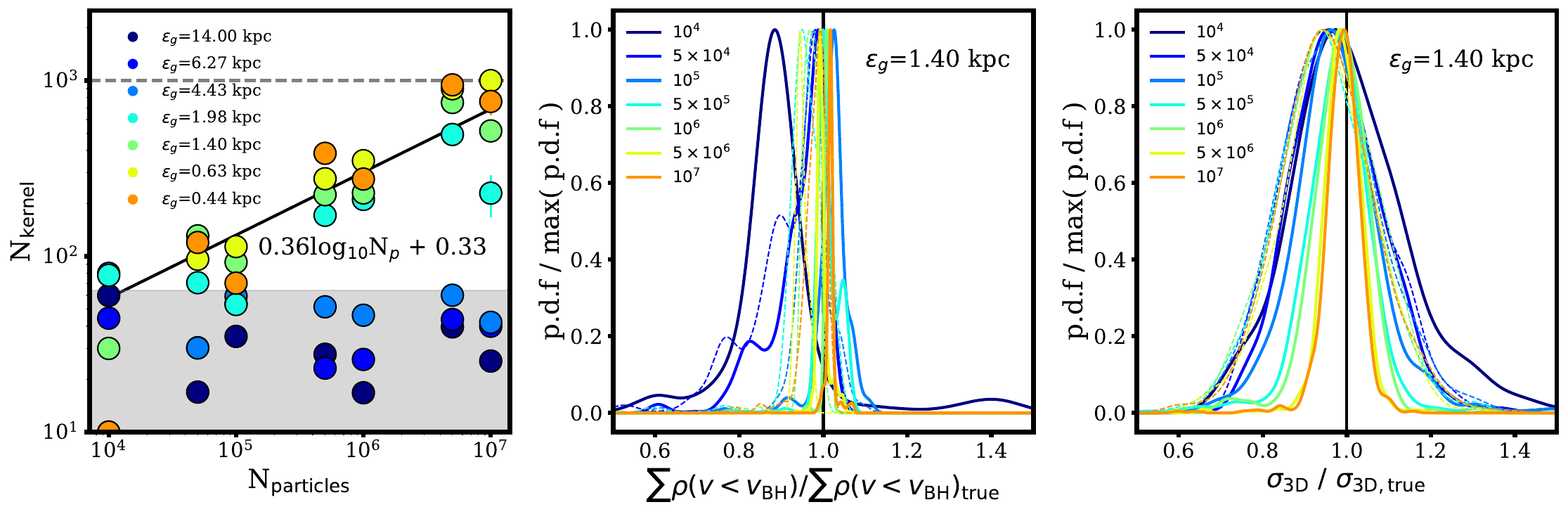}
    \caption{\textit{Left:} the number of particles in the kernel required to obtain an accurate estimate of $\sum \rho(v<v_{\rm BH})$ as a function of the number of particles in the halo. The points are coloured by the size of the gravitational softening from largest (blue) to highest (orange). The error bars are shown in each case, but are in many cases smaller than the point size. The black line is the best linear fit to the relation. The maximum number of particles in the kernel is limited to 1000 (grey dashed line) and the minimum is capped at 64 particles (grey shaded region). \textit{Middle:} histograms with solid lines show the accuracy of the cumulative density encountered by the black hole in halos with varying mass resolution with a fixed softening of 1.4~kpc. The densities were computed using our optimal-kernel-size method and compared to the prediction from the Hernquist distribution function. Thin dashed lines show the estimate using a kernel defined by 64 nearest neighbouring particles as in \citet{tremmel}. \textit{Right:} the accuracy of the estimate of the 3D velocity dispersion with a fixed softening of 1.4~kpc. Solid lines show the results using our optimal kernel size, and dashed lines show the results for a kernel with 64 particles.}
    \label{fig8}
\end{figure*}

As pointed out previously, our calibration procedure is based on the assumption that the density $\rho(v<v_{\rm BH})$ is known exactly and so is the velocity dispersion, $\sigma_{\rm 3D}$ (which determines $b_{\rm min}$). However, as we have seen in Fig.~\ref{fig3}, the density estimate is likely to be noisy. In our method, we should thus aim to minimize this noise while simultaneously maintaining the accuracy of the density and velocity dispersion estimates. 

In order to obtain an optimal kernel size at each resolution level, we first take a fixed black hole orbit from the highest-resolution run shown in the left panel of Fig.~\ref{fig2} ($10^7$ particles, $M_{\rm BH}/M_{\rm DM} = 1000$), defined by 3D positions and velocities, ($\mathbf{r}_{\rm BH}(t)$, $\mathbf{v}_{\rm BH}(t)$). We then compare the values of $\rho(v<v_{\rm BH})$ and $\sigma_{\rm 3D}$ measured in dark-matter-only simulations for each ($\mathbf{r}_{\rm BH}(t)$, $\mathbf{v}_{\rm BH}(t)$) at each resolution level to those predicted by the Hernquist distribution function\footnote{Note that for the distribution function prediction, we use the velocity of the black hole relative to particles within the kernel.}. Note, however, that we opt to compare the \textit{cumulative} density encountered by the black hole along its trajectory, $\sum \rho = \sum_{t = 0}^{t = {\rm current}} \rho(v<v_{\rm BH})$,  with the analytical prediction. This is because a density estimate that is noisy but on average accurate may still sink the black hole on the correct timescale, compared to a smooth estimate with a smaller sum of individual density residuals, but a position-dependent bias. We additionally estimate the errors for each $\sum \rho$ and $\sigma_{\rm 3D}$ by incrementally rotating the trajectory of the black hole in 3 dimensions, such that we have a distribution of $\sum \rho$ and $\sigma_{\rm 3D}$ at fixed ($d_{\rm BH}(t)$, $\mathbf{v}_{\rm BH}(t)$), where $d_{\rm BH}$ is the distance of the black hole from the centre of the halo.

We note that as the black hole approaches the halo centre and encounters steeper density gradients, a large kernel size can bias the density estimate.  For this reason, we define a ``maximum resolved density''\footnote{This expression is obtained from the condition $\rho_{\rm max} = 64\, m_{\rm DM} / (\frac{4}{3} \pi \epsilon_g^3)$, and setting $\epsilon_g$ as in Eq.~(\ref{power_eq}).}, $\rho_{\rm max} \approx \rho_{\rm 200,crit} \sqrt{N_{\rm p}} $. We found that this density approximates well the transition in the density profile where smaller kernel sizes are more advantageous than larger ones. We then sample the kernel sizes in the range 10 to 1000 particles using {\small EMCEE} for the trajectory of the black hole in regions of the halo where the local density is smaller than $\rho_{\rm max}$, computing the log-likelihood:
\begin{equation}
\begin{split}
\ln L_k  = -\frac{1}{2} \sum_{t=0}^{t = { t(\rho_{\rm max})}} \left( \frac{\sum \rho(N_k) - \sum \rho_{\rm analytical}}{\sigma_{\sum\rho (N_k)}} \right.~+ \\ \left.\frac{\sigma_{\rm 3D}(N_k) - \sigma_{\rm 3D, analytical}}{\sigma_{\sigma_{\rm 3D} (N_k)}} \right),
\end{split}
\end{equation}
where $\sigma_{\sum \rho}$ and $\sigma_{\sigma_3D}$ are the errors on the estimates. 

The results can be seen on the left panel of Fig.~\ref{fig8}, through which we derive a relation: 
\begin{equation}
\log_{10} N_{\rm k} = 
\begin{cases}
{\rm min}(0.36\log_{10} N_{\rm p} + 0.33,~3)  & \text{if} ~ \rho<\rho_{\rm max}, \\
 \log_{10} 64 & \text{if} ~ \rho > \rho_{\rm max}, \\
 \log_{10} 64 & \text{if} ~ \epsilon_g/a_{\rm HQ} > 0.03,
\end{cases}
\label{kernel_eq}
\end{equation}
where $N_{\rm p}$ is the number of particles in the halo, and $N_{\rm k}$ is the optimal kernel size. We chose to limit the number of particles in the kernel to a minimum of 64 and a maximum of 1000. The middle panel of Fig.~\ref{fig8} shows how the accuracy of estimates of $\sum \rho(v<v_{\rm BH})$ using our optimal kernel size compare to the values predicted by the Hernquist distribution function. The thin lines shows the result for a fixed kernel size of 64 particles as in \citet{tremmel}, where the benefit of our approach at higher resolution is particularly evident. In practice, as a typical treatment for black hole seeding in cosmological simulations involves identifying the mass of the host halo, $N_{\rm p}$ can be approximated simply as $M_{\rm halo} / m_{\rm DM}$.

\subsection{A treatment for low-mass black holes}
\label{sec_low_mass}

\begin{figure*}
    \centering
    \includegraphics[width=2\columnwidth]{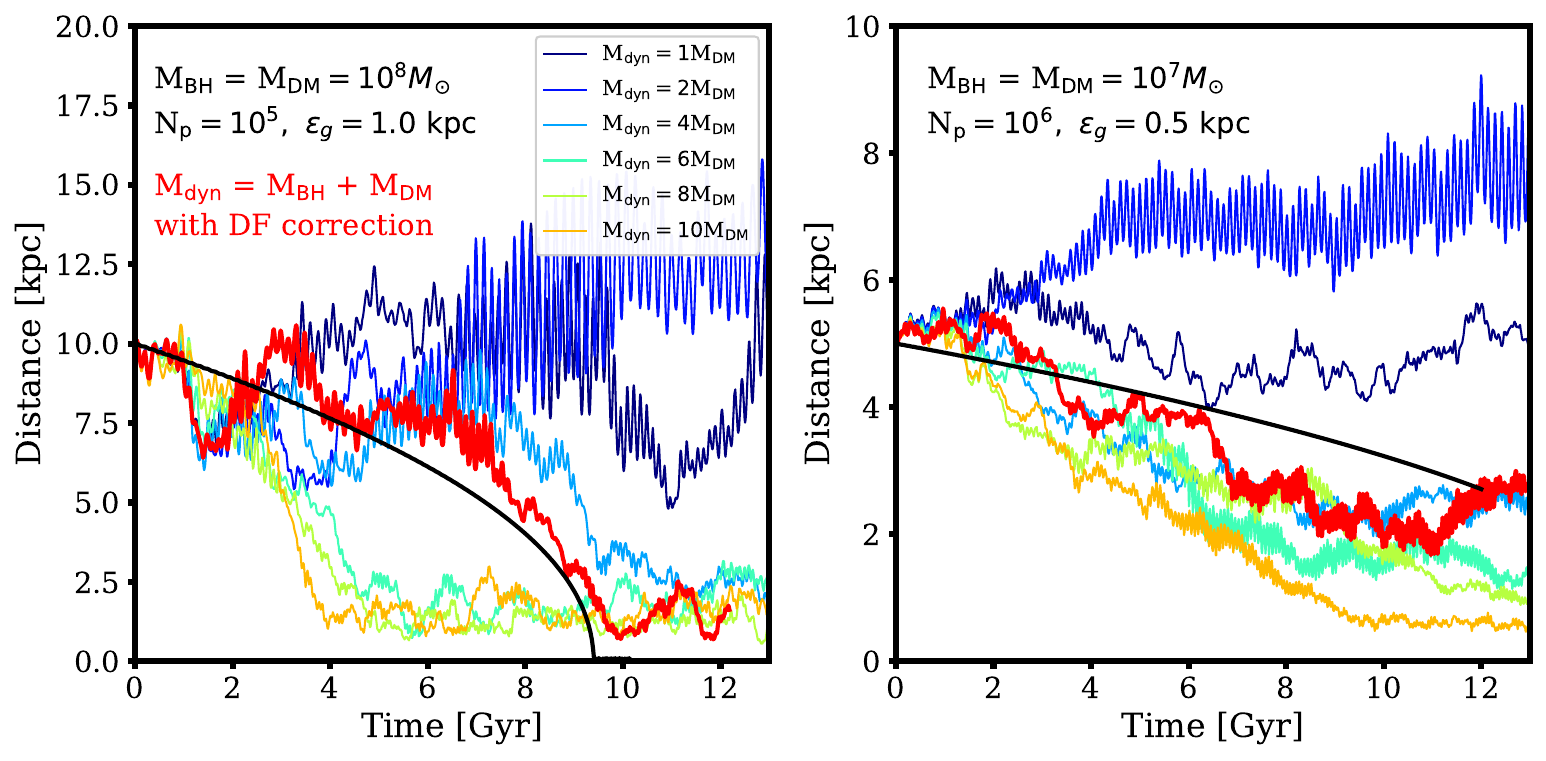}
    \caption{\textit{Left:} the effect of increasing the dynamical mass of black holes on their orbital trajectories. In black we show the analytical prediction for the orbit of a $10^{8}\, {\rm M}_{\odot}$ black hole in a $10^{13}\, {\rm M}_{\odot}$ halo, starting on a circular orbit at 10~kpc. The initial orbit with $M_{\rm BH}/M_{\rm DM} = 1$ is shown with a thin dark blue line. The dynamical mass of the black hole is then scaled by a factor $k$ in the range 1 to 10 times the dark matter particle mass, resulting in faster black hole sinking. No dynamical friction correction is applied. The thick red line shows the result of setting the dynamical mass of the black hole to $M_{\rm dyn} = M_{\rm BH} + M_{\rm DM}$ and applying the correction for unresolved dynamical friction via Eqn.~(\ref{mymodel}). \textit{Right:} same as the left panel, but for a $10^7\,{\rm M}_{\odot}$ black hole starting at 5~kpc. The black hole trajectories have been smoothed over 0.1~Gyr for clarity.}
    \label{fig9}
\end{figure*}

Finally, we would like our method to be able to deal with the case of low-mass black holes (i.e.~black holes with mass less or equal to the dark matter particle mass). As we have already seen, for the case $M_{\rm BH}/M_{\rm DM} = 1$ the black hole behaves in a similar way as the dark matter and does not sink. For $M_{\rm BH}/M_{\rm DM} < 1$, we furthermore expect to see effects of mass segregation \citep{ludlow}, where lower-mass particles are scattered to larger radii by higher-mass ones. 

Within the framework of our model, we can make low-mass black holes sink either by increasing their dynamical mass, increasing the magnitude of the Coulomb logarithm, or fully accounting for the heating due to stochasticity for a given background particle mass. 

While increasing the dynamical mass of the black hole may be in part physically motivated as many black holes are observed to live inside star clusters (see \citealt{Kormendy2013, neumayer_review} and references therein), we do not at present have a subgrid prescription for the formation of these clusters. It is also unclear to what extent would these clusters enhance the dynamical mass of the black hole, with observations suggesting star clusters both more and less massive than the black holes they host, and whether this boost in mass would guarantee sinking at given mass resolution.

In Fig~\ref{fig9}, in black, we show the circular orbit of a 10$^8\,{\rm M}_{\odot}$ and a $10^7\,{\rm M}_{\odot}$ black hole in a $10^{13}\,{\rm M}_{\odot}$ halo. In both cases, $M_{\rm BH}/M_{\rm DM} = 1$. Coloured lines show the orbit of a black hole as we increase its dynamical mass by a factor $k$ between 1 and 10. It can be seen that the mass ratios $M_{\rm dyn}/M_{\rm DM} < 4$  are often not enough to guarantee sinking of the black hole without added dynamical friction \citep{bellovary}. On the other hand,  $M_{\rm dyn}/M_{\rm DM} \gtrsim 4$ are often sufficient to sink the black hole. However, while increasing the dynamical mass may be physically justified if the black hole is embedded in a more massive star cluster, it may also lead to undesirable effects such as artificial scattering of background particles by the black hole \citep{wurster}. In any case, Fig.~\ref{fig9} demonstrates that there is no particular dynamical mass scaling that would reproduce the black hole orbit for all initial conditions. 

Our calibration method, described in Section~\ref{themethod} with results shown in Figs.~\ref{fig6} and~\ref{fig7}, suggests a strategy for increasing the magnitude of the Coulomb logarithm to compensate for the stochasticity, as our derived optimal values of $\eta$ for each $M_{\rm BH}/M_{\rm DM}$ marginalise over the impact of noise. However, $\eta$ is not well constrained for low mass ratios and is close to a flat distribution, as shown in Fig.~\ref{fig5}, due to stochastic forces having comparable magnitude to that of dynamical friction i.e at any given position the force that the black hole experiences can deviate significantly from that due to a smooth Hernquist halo. We thus conclude it would be more advantageous to derive an estimator for the overall effect of stochasticity from first principles.

As discussed in Section~\ref{diff_coeff}, while we can obtain an estimate of the magnitude of the stochastic acceleration, we cannot at any given point determine the net direction of the stochastic kicks. In some cases, stochastic effects can ultimately lead to a \textit{more} rapid black hole sinking (right panel of Fig.~\ref{fig1}). However, as the effect of stochasticity is to increase the kinetic energy of the black hole, we can derive an energy-loss term to counteract this.  

We thus will assume that the \textit{overall} effect of the second-order diffusion coefficients, working against the effect of dynamical friction, has the form:
\begin{equation}
a_{\rm noise} = \frac{4 \pi G^2 \rho(v < v_{\rm BH}) M_{\rm DM}}{v^3} \ln\left(\frac{R}{\eta \epsilon_g} \right) {\bf v}  {~~~\rm if~} R>\eta\epsilon_g,
\label{stoch_correction}
\end{equation}
where $M_{\rm DM}$ is the background particle mass, which brings the noise correction to the right order of magnitude (see Appendix A). Since particle scattering below $\eta \epsilon_g$ is unresolved, only impact parameters above $\eta \epsilon_g$ contribute, giving the Coulomb logarithm $\ln \Lambda = \frac{R}{\eta \epsilon_g}$. We note that this is different to applying the term ${\bf a_{\rm df,m_{*}}}$ as in the work of \citet{damiano} -- their additional ``cooling'' force accounts for \textit{unresolved} dynamical friction from the mass of background particles (assuming their masses are the true masses of background stars and dark matter), while Eq.~(\ref{stoch_correction}) attempts to negate the \textit{resolved} heating of the black hole by massive background particles.

We note that the application of the noise correction is, in principle, equivalent to setting the dynamical mass of the black hole in the gravity solver to
\begin{equation}
    M_{\rm dyn} = M_{\rm BH} + M_{\rm DM},
    \label{dyn_mass}
\end{equation}
while the correction for the unresolved part of dynamical friction is still computed for a black hole mass $M_{\rm BH}$. The mass rescaling approach may also be more advantageous in practice, as computing the distance to the centre of the halo at each time step entails a higher computational cost. Additionally, the definition $b_{\rm max} = R_{\rm BH}$ comes from the assumption of a Hernquist halo \citep{just_khan} and can change for other density distributions. The exact definition of $b_{\rm max}$ is irrelevant in mass boosting. Nevertheless, in practice, we find that directly applying the noise term will sink the black holes more aggressively and guarantees sinking, while increasing the dynamical mass as per Eq.~(\ref{dyn_mass}) reproduces the orbital path of the black holes more accurately on average, but in some cases can result in black hole sinking that is substantially delayed or even suppressed. 

The result of the application of this stochasticity correction through a rescaling of the dynamical mass, together with the correction for unresolved dynamical friction, is shown with a thick red line in Fig~\ref{fig9}. In neither case does the correction fully reproduce the true trajectory of the black hole, however, in both cases the black holes sink to the centre of the galaxy without the need for a substantial increase of their dynamical mass.

\subsection{Black holes in the regime $1 \leq M_{\rm BH}/M_{\rm DM} \leq 5$}
\label{test_corr}

\begin{figure}
    \centering
    \includegraphics[width=\columnwidth]{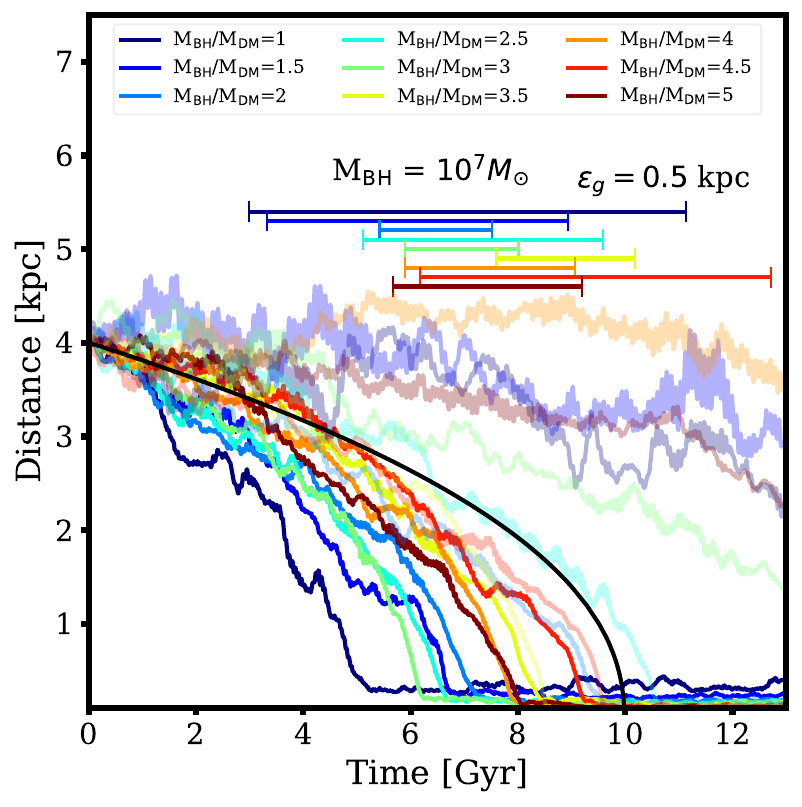}
    \caption{The results of applying our dynamical friction correction, coupled with the correction for stochasticity, to a $10^7\, {\rm M}_{\odot}$ black hole in a $10^{13}\, {\rm M}_{\odot}$ halo starting at 4~kpc. Darker lines show median black hole trajectories for different $M_{\rm BH}/M_{\rm DM}$ ratios as the halo mass resolution changes. The error bars show the range of sinking times of black holes from the sample of 6 initial conditions used to compute the median trajectories. The lighter lines show examples of trajectories where only a dynamical friction correction is applied, i.e.~no stochasticity correction. The black hole trajectories have been smoothed over 0.15~Gyr for clarity. }
    \label{fig10}
\end{figure}

We now explore how our stochasticity correction behaves in the regime $1 \leq M_{\rm BH}/M_{\rm DM} \leq 5$. This is shown in Fig.~\ref{fig10}, where for a fixed halo and black hole mass, we gradually increase the mass resolution such that  $M_{\rm BH}/M_{\rm DM}$ varies from 1 to 5. The dark lines show the median trajectory of 6 black hole seeds for each mass ratio. We apply the stochasticity correction via Eq.~(\ref{stoch_correction}). The correction clearly tends, on average, to sink the black holes too fast, with the effect being strongest for lower $M_{\rm BH}/M_{\rm DM}$.  The error bars range from the earliest to the latest sinking times of the black holes, where the mass ratio appears to set the earliest black hole sinking time, while the maximum sinking time does not seem to depend strongly on the resolution. The light lines show individual realisations of black holes at each mass ratio where only the dynamical friction correction is applied. In some cases, this faithfully recovers the black hole orbit ($M_{\rm BH}/M_{\rm DM} = 2$, $2.5$, and $4.5$), but for others the black holes do not sink within the age of the Universe. Interestingly, this does not seem to depend strongly on the mass ratio in this regime (some black holes with $M_{\rm BH}/M_{\rm DM} = 4$ and~5~do not sink even with a dynamical friction correction). We conclude that a stochasticity correction is required to ensure that a black hole sinks to the centre of a simulated halo, however the inherent randomness of this process makes it difficult to recover black hole orbits accurately for any individual case in the regime $M_{\rm BH}/M_{\rm DM} \leq 5$. 

We have repeated this test with rescaling the dynamical mass of the black hole to $M_{\rm BH} + M_{\rm DM}$ instead of applying the stochasticity correction directly. We found that up to $M_{\rm BH}/M_{\rm DM} \leq 5$, 10~per-cent of the initial conditions lead to substantailly delayed sinking of the black holes (the black holes do not sink before 13~Gyr) and 13~per~cent of initial conditions show no or little sign of sinking over 13 Gyr, while the median sinking times for each $M_{\rm BH}/M_{\rm DM}$ lie between 8 and 12~Gyr. Directly applying the stochastic correction via Eqn.~(\ref{stoch_correction}), on the other hand, at all times guarantees the sinking of black holes. 

We believe the discrepancy between the two arises for two reasons. One is the scattering that the black hole of higher dynamical mass induces, which can alter the properties of the surrounding medium. Another is that, when the correction is applied directly the net force is always exactly opposite to the direction of motion of the black hole, maximising the reduction in kinetic energy.

In conclusion, in the regime $1 \leq M_{\rm BH}/M_{\rm DM} \leq 5$, the predictions for individual black hole trajectories become highly uncertain when using a dynamical friction subgrid model such as the one described in this work. To guarantee black hole sinking, one needs to account for the effects of stochasticity e.g. via Eq.~(\ref{stoch_correction}). The dynamical mass rescaling as in Eq.~(\ref{dyn_mass}) ($M_{\rm dyn} = M_{\rm BH} + M_{\rm DM}$) works well on average, but in $\sim 13$~per~cent of cases this mass scaling is too weak to sink the black hole. Increasing the dynamical mass to $M_{\rm dyn} \geq 4 M_{\rm DM}$ would guarantee sinking, but at the price of lost accuracy in the orbital trajectory. We also note that the regime of $M_{\rm BH}/M_{\rm DM} < 1$ has not been explored in this work. As such, in the framework of this model, we recommend that the black holes are seeded at least with the mass of the dark matter particles.

\section{Applying the model}
\label{summary_sec}

\begin{figure*}
 \centering
    \includegraphics[width=2\columnwidth]{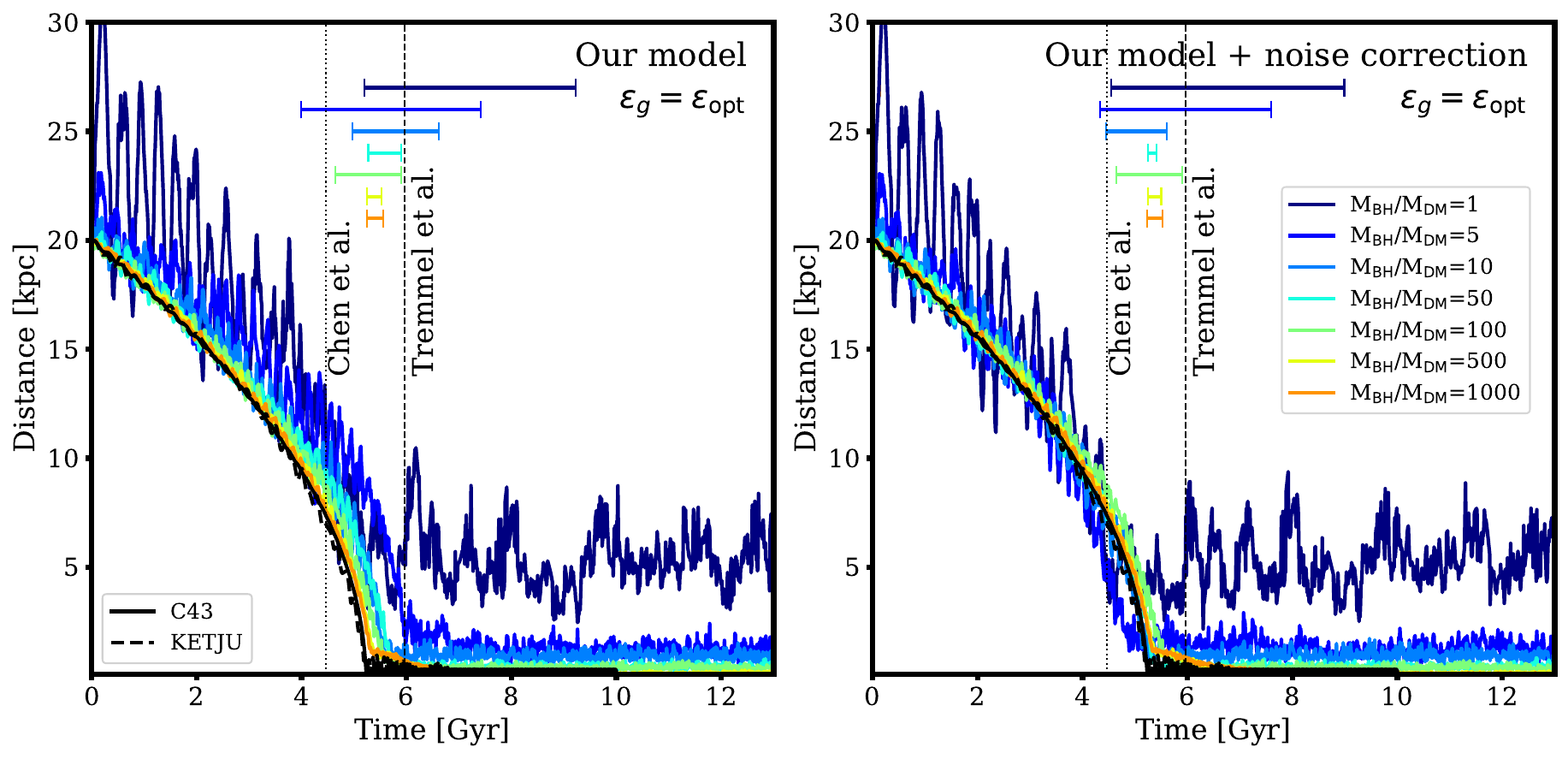}   \\
    \includegraphics[width=2\columnwidth]{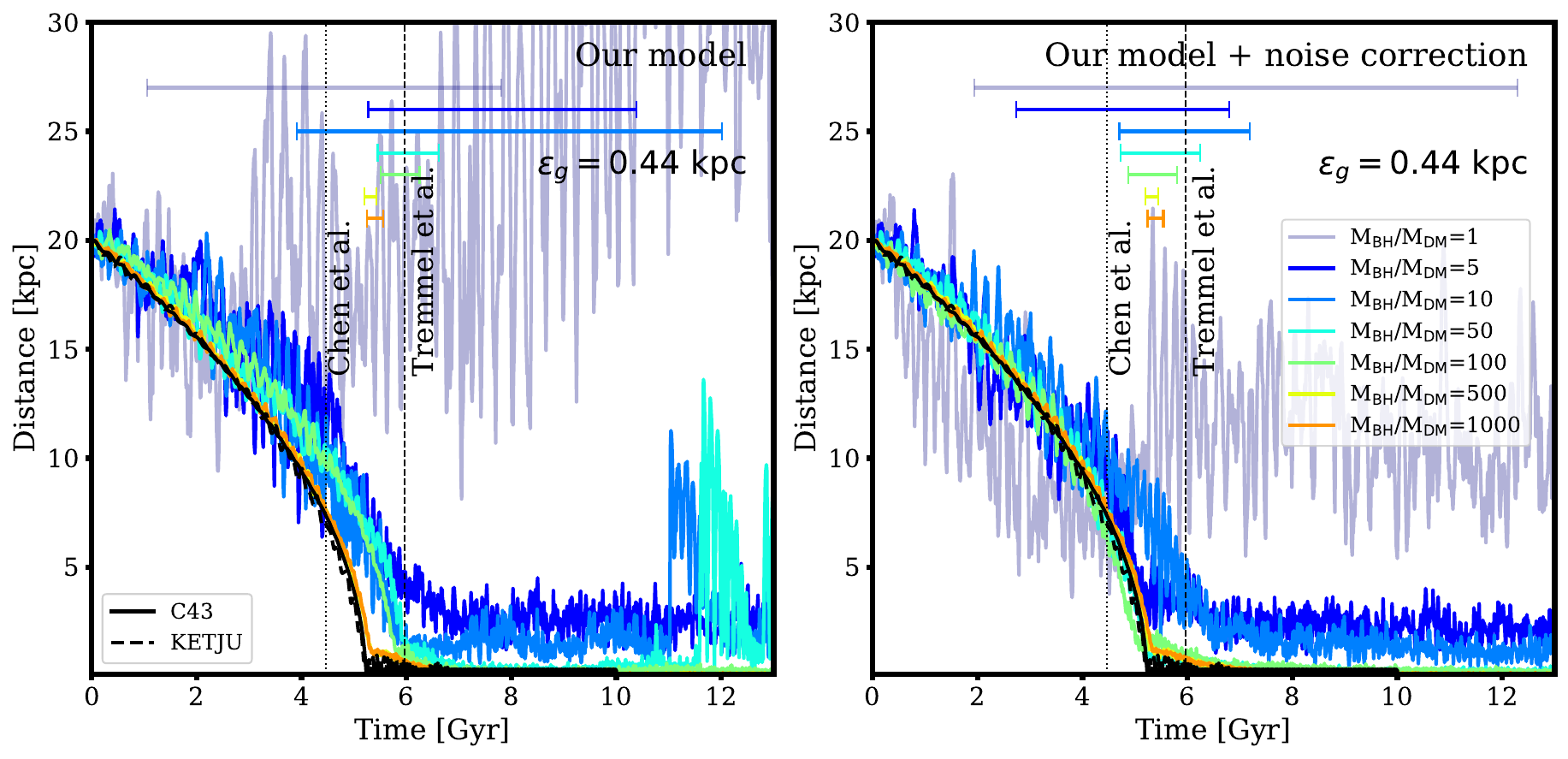}

    \caption{\textit{Top:} the median trajectories of $10^9\,{\rm M}_{\odot}$ black holes in a $10^{13}\,{\rm M}_{\odot}$ halo computed using our calibrated dynamical friction model. The results are shown for 7 mass resolutions and their corresponding optimal softening values (see Table~\ref{table1}). The grey dotted and dashed lines show the sinking time of the black hole at the highest background particle mass resolution using the \citet{chen} and \citet{tremmel} models, respectively. The error bars show the range of sinking times computed from 6 simulations at each resolution with differing random seeds in the initial conditions. The sinking time is defined as the time when the black hole first comes one softening length away from the centre of the halo. The left and right panels show the results with and without the application of the noise correction, respectively. \textit{Bottom:} same as the top panel, but using a fixed softening of $\epsilon_g = 0.44$~kpc.}
    \label{fig11}
\end{figure*}

In this Section, we briefly summarise the necessary components of our model and the requirements for its implementation in a cosmological simulation. We then apply the model to a number of test cases.

\subsection{Summary of the model}

\begin{enumerate}
    \item When the black hole is seeded in a halo of given mass, the number of particles in the halo is approximated as $N_p=M_{\rm halo}/M_{\rm DM}$ and the ``maximum resolved density'' as $\rho_{\rm max}=\rho_{\rm 200,crit}\sqrt{N_{p}}$. The kernel size for estimating the local density and velocity dispersion for each black hole can then be determined through Eq.~(\ref{kernel_eq}). This value can be updated when the halo finder is applied again \footnote{Since many galaxy formation models employ an on-the-fly halo finder to seed the black holes (e.g IllustrisTNG, MAGNETICUM), this additional step should not substantially increase the computational cost required by our model.}.

    \item The scale radius of each halo, $a_{\rm HQ}$, may be estimated via, e.g., the mass-concentration relation in $\Lambda$CDM \citep{ludlow_mass_conc} and eq.~2 of \citet{sdmh2} \footnote{Note that the relative flatness of the relation in Eq.~(\ref{optimal_eta_scaled}) implies low sensitivity to the scale length, the determination of which may be impacted by ongoing mergers, or by the presence of multiple populations of background particles with differing $a_{\rm HQ}$.}. For a given gravitational softening of the black hole, $\epsilon_g$, Eq.~(\ref{optimal_eta_scaled}) is used to determine the maximum unresolved impact parameter, $b_{\rm max, unres} = \eta (\epsilon_g) \epsilon_g$.  We emphasise that the relation in Eq.~(\ref{optimal_eta_scaled}) is only applicable to codes that use the same spline softening as in  {\small GADGET} and {\small AREPO} \citep{arepo,gadget4}.
   
    \item The black hole is assigned a dynamical mass $M_{\rm dyn} = M_{\rm BH} + M_{\rm DM}$, which is the mass used in the gravity solver. The dynamical mass should be updated as long as M$_{\rm BH}/M_{\rm DM} \leq 100$, subject to the following constraints:
    \begin{itemize}
    \item The black hole dynamical mass should always be at least twice the dark matter particle mass\footnote{Based on the range of $M_{\rm BH}/M_{\rm DM}$ investigated in this work.}.
    \item As an alternative to dynamical mass rescaling, one may wish to apply the noise correction directly to the black hole, in which case one needs to estimate at each timestep the distance of the black hole from the centre of the halo\footnote{While this is based on tests using isolated halos, we note that, for large $R$, inaccuracy in the distance estimate does not substantially alter the value of the Coulomb logarithm.}, $R$.
    \end{itemize}

    \item At each timestep, the local density of particles moving slower than the black hole, $\rho (v < v_{\rm BH})$, and the 3D velocity dispersion, $\sigma_{\rm 3D}$, are computed from the properties of the nearest $N_{k}$ particles. The minimum impact parameter is computed as $b_{\rm min} = GM_{\rm BH}/(v^2_{\rm BH} + 2/3\, \sigma^{2}_{\rm 3D})$, where the black hole velocity is relative to the mean velocity of the surrounding particles.
     In case of multiple populations (e.g. stars and dark matter), the properties are computed separately for each population and so is the resultant dynamical friction. $N_k$ should also be computed for different populations. If a negligible density of particles is present in a given population, the dynamical friction contribution may be ignored.
    
    \item The correction for unresolved dynamical friction is applied using Eq.~(\ref{mymodel}), with the Coulomb logarithm computed as $\ln \Lambda = b_{\rm max, unres}/b_{\rm min}$. The correction is applied as long as $b_{\rm min} < b_{\rm max, unres}$. If the dynamical mass of the black hole has not been rescaled, a stochasticity correction should be applied using Eq.~(\ref{stoch_correction}). 
\end{enumerate}

\subsection{Validation}

We will now demonstrate how our method performs in a variety of setups. In particular, we will test our fiducial circular orbit setup, a more eccentric orbit and the test case presented in \citet{tremmel}.

\subsubsection{Fiducial set-up}

In Fig.~\ref{fig11}, we show the median black hole trajectories generated from 6 random realisations of the Hernquist halo at 7 different mass resolutions. The top panel shows simulations with the optimal \citet{power} softening, while the bottom panel shows the results with a fixed softening of 0.44~kpc. The left and right panels give the results without and with the stochasticity correction, respectively. We compare our results to those using the \citet{tremmel} and \citet{chen} models for the highest-resolution case ($N_{\rm p} = 10^7$, $M_{\rm BH}/M_{\rm DM}=1000$).

It is evident that our calibrated formula reproduces well the sinking time of the black holes from the analytical and {\small KETJU} predictions. With $M_{\rm BH}/M_{\rm DM} \geq 100$ the sinking times are well converged, while at lower mass ratios a stochasticity correction is clearly required, as deviations from the prediction become apparent even at large radii. The importance of optimal softening choice is especially evident in the lower panel of Fig.~\ref{fig11}, where the softening of $\epsilon_g=0.44\,{\rm kpc}$ is clearly too small for halos with $N_{\rm p} \le 10^5\, {\rm M}_{\odot}$, leading to strong stochastic effects. Black holes with $M_{\rm BH}/M_{\rm DM} = 1$, for instance, tend to get ejected away from the halo centre, even with application of the unresolved dynamical friction force. 

The fluctuations seen in the median black hole orbit at 11~Gyr for cases with $M_{\rm BH}/M_{\rm DM}=10$ and $M_{\rm BH}/M_{\rm DM}=50$ correspond to individual black holes that only arrive to the centre of the halo at this time. In one case with $M_{\rm BH}/M_{\rm DM} = 10$, the black hole sinks to the centre of the halo at 6 Gyr, but then gets ejected from the centre before sinking again (a rise in the median profile between 6 and 8 Gyr). The stochasticity correction (Eq.~\ref{stoch_correction}) avoids these effects (bottom right panel). For the case with $10^4$ particles, a constant density core on the scale of $\sim7\, {\rm kpc}$ develops, leading to the black hole stalling around this radius and additionally making it difficult to define the halo density centre.

\subsubsection{An eccentric orbit}

\begin{figure}
 \centering
    \includegraphics[width=\columnwidth]{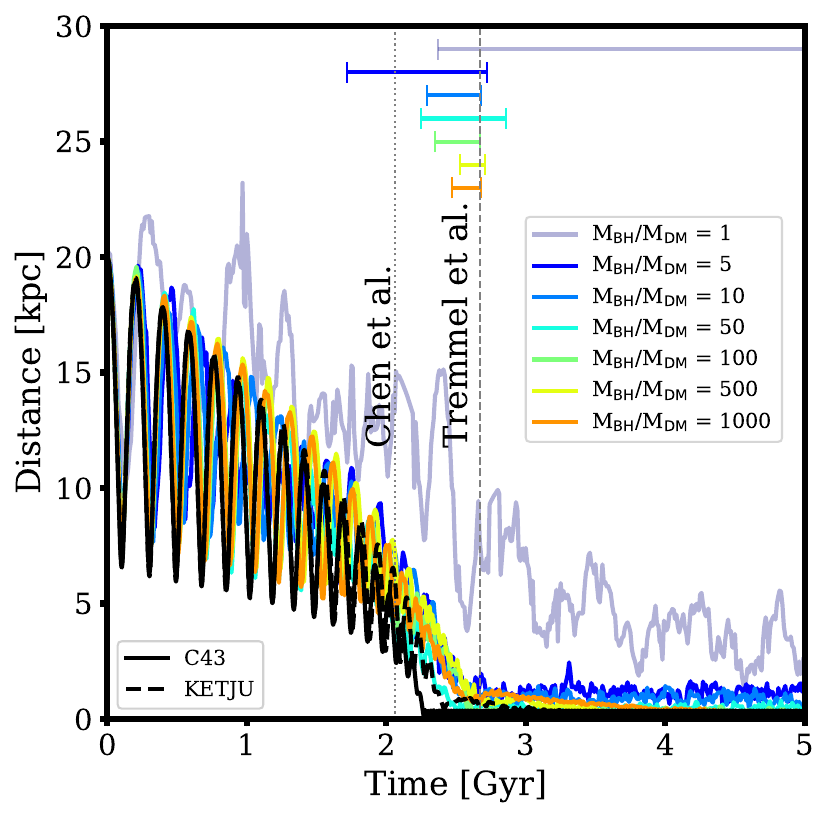}
    \caption{The median trajectories of $10^9\, {\rm M}_{\odot}$ black holes set on an eccentric orbit in a $10^{13}\, {\rm M}_{\odot}$ halo. The median trajectories are computed from 6 random seeds at each resolution level (same as Fig.~\ref{fig11}). The error bars show the range between the earliest and the latest black hole sinking time. The black hole dynamics were computed using our correction for the unresolved dynamical friction and the correction for stochasticity. The grey dotted and dashed lines show the sinking time for $M_{\rm BH} / M_{\rm DM}=1000$ predicted by the \citet{chen} and \citet{tremmel} models, respectively. }
    \label{fig12}
\end{figure}

We now test our method on a more eccentric black hole orbit. The $10^9\,{\rm M}_{\odot}$ black hole in a $10^{13}\,{\rm M}_{\odot}$ halo has a starting velocity $v_{\rm BH}=0.5\,v_{\rm circ}$ at 20~kpc, resulting in a sinking time of $2.3-2.4\,{\rm Gyr}$. The results of the test are shown in Fig.~\ref{fig12}. This time, we see a mild difference in sinking time ($\sim 0.15$~Gyr) between the analytical prediction and {\small KETJU}. This likely stems from the limited resolution at the centre of the halo. The thick lines show the median black hole trajectories, where we see that the dynamical friction and stochastic correction are in this case slightly too weak, with black hole sinking time delayed on average by $\sim0.2$~Gyr. This may seem contradictory to our findings in Figs.~\ref{fig10}~and~\ref{fig11}, but we emphasize that both of those represent extreme cases with long black hole sinking timescales over which stochastic effects become obvious.

\subsubsection{An example from \citet{tremmel}}

\begin{figure}
 \centering
    \includegraphics[width=\columnwidth]{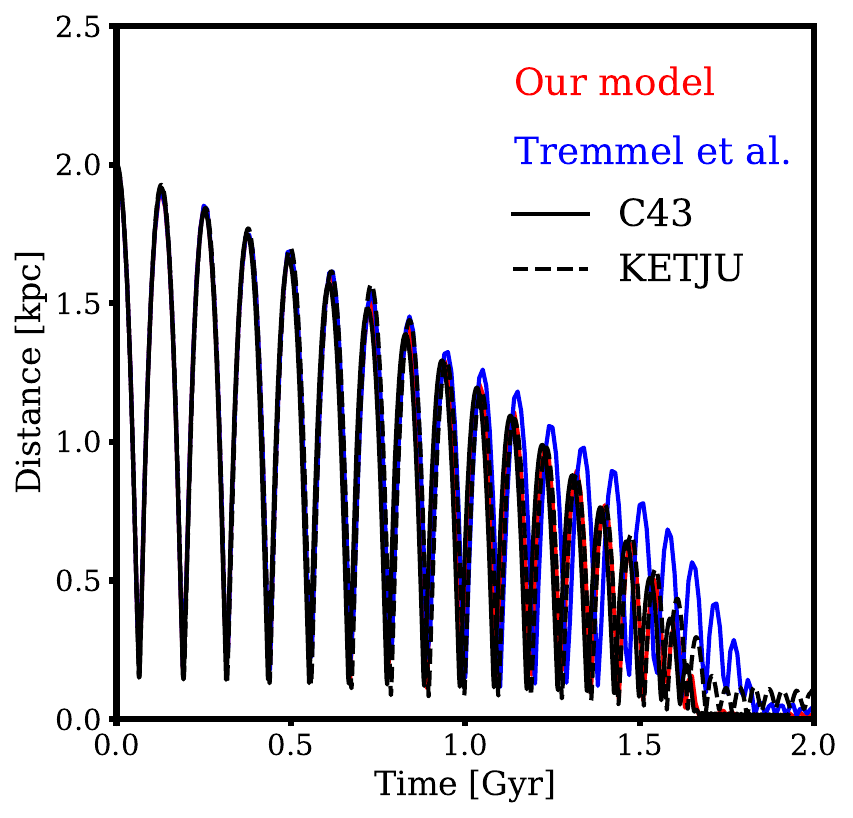}   
    \caption{The trajectory of a $10^6\, {\rm M}_{\odot}$ black hole in a $2\times10^{11}\,{\rm M}_{\odot}$ halo from the {\sc HighRes} example presented in \citet{tremmel}. The red line shows the result of applying our model, while the blue line is the result of the \citet{tremmel} model. The black solid and dashed lines are the analytical and {\small KETJU} predictions, respectively. All simulations were carried out with $M_{\rm BH}/M_{\rm DM} = 376$.}
    \label{fig13}
\end{figure}

Finally, we apply our model to the initial conditions similar to the {\sc HighRes} test case presented in \citet{tremmel}. This includes a $M_{200}=2\times10^{11}\, {\rm M}_{\odot}$ halo with concentration $c = 4.5$ and $R_{200}=115$~kpc, in which a black hole with mass $10^6\, {\rm M}_{\odot}$ starts off at 2~kpc, with a velocity $v=0.1\,v_{\rm circ}$. The dark matter particle mass in our simulations is $2.7\times10^{3}\, {\rm M}_{\odot}$, resulting in $M_{\rm BH}/M_{\rm DM}\approx 380$. The gravitational softening is set to $\epsilon_g = 77$~pc, as in \citet{tremmel}. In Fig.~\ref{fig13}, the red line shows the black hole trajectory using our dynamical friction model, and the blue line shows the trajectory of the black hole using the \citet{tremmel} model. The sinking time prediction for the analytical model and {\small KETJU} differs by $\sim$0.05~Gyr. This could be the result of minor stochastic effects we expect for mass ratios $100 < M_{\rm BH}/M_{\rm DM} < 500$ (see Fig.~\ref{fig4}).  We also note that this was the most computationally expensive of our simulations in this work, as the accuracy of the time integration in the analytical model (and consequently the simulations with our model and the model of \citealt{tremmel}) had to be increased to 0.1~per~cent in order to accurately capture the eccentric orbit of the black hole, as confirmed by convergence tests. Our model follows the analytical prediction and that of {\small KETJU} almost exactly, while the model of \citet{tremmel} results in the black hole sinking time delayed by almost $0.2$~Gyr.

\section{Conclusions}
\label{conclusion_sec}

Dynamical friction is one of the main factors governing black hole motion, driving black holes to the centres of galaxies where they can accrete and grow. It is therefore essential to model this process accurately in cosmological hydrodynamics simulations. Unfortunately, the spatial and mass resolution required to capture the full extent of dynamical friction comes at an extremely high computational cost, especially for large cosmological volumes. In particular, the use of the gravitational softening limits the scattering of local particles by the black hole, which would otherwise contribute a significant fraction of the total dynamical friction force. Coarse-grained representation of background stars and dark matter also introduces artificial heating of the black hole orbit. The latter effect is particularly problematic in cases where the black hole particle mass is of similar magnitude to that of the background stars and dark matter.

A number of numerical models aimed at compensating for the unresolved part of the dynamical friction force have been introduced in the literature \citep{tremmel, pfister, chen, damiano}. These models assume that the effect of dynamical friction in an $N$-body simulation is well-described by the Chandrasekhar formula \citep{chandrasekhar1943} in which the minimum impact parameter of the Coulomb logarithm, $\ln\Lambda= \frac{b_{\rm max}}{b_{\rm min}}$, is determined by the spatial resolution of the simulation ($b_{\rm min} \approx \epsilon_g$). They then apply a correction for the unresolved part of dynamical friction, with  $\ln\Lambda_{\rm unresolved} =\frac{b_{\rm max, unres}}{b_{90}} \approx \frac{\epsilon_g}{b_{90}}$, where $b_{90}$ is the impact parameter for a 90-degree deflection. Such a correction has been shown to effectively sink black holes to galactic centres on roughly correct timescales. In the case of low-mass black holes ($M_{\rm BH} \sim M_{\rm DM}$), these models artificially boost the dynamical mass of the black hole or increase the strength of dynamical friction to overpower the effects of the numerical noise, with the main goal of keeping low-mass black holes near the centres of their host galaxies. 

Nevertheless, it has remained unclear what is the exact spatial scale up to which dynamical friction is unresolved in $N$-body simulations and how this scale relates to the choice of gravitational softening. The effect of numerical noise on black hole motion has also been largely ignored in the literature. These are the issues we have aimed to address in this work. By comparing the dynamics of black holes in simulations with varying spatial and mass resolution to high-resolution simulations run with the code {\small KETJU}, in which the gravitational interaction of black holes are not softened, we were able to calibrate a subgrid model based on the Chandrasekhar formula to accurately follow orbits of black holes and their sinking to halo centres. In particular:

\begin{enumerate}
 \item{We have shown that the definition of the impact parameter for a 90-degree deflection, $b_{\rm min}$, is important. The definition $b_{\rm min}=GM_{\rm BH}/(v_{\rm BH}^2 + 2/3 \,\sigma_{3D}^2)$, taking into account random motions in the typical velocity of an encounter, ensures that $b_{\rm min}$ is generally converged when changing mass resolution \citep{just_khan}. The definition $b_{\rm min}=GM_{\rm BH}/v_{\rm BH}^2$, combined with $b_{\rm max} =\epsilon_g$, as in \citet{tremmel}, tends to sink the black holes too slowly.}

 \item{The maximum unresolved impact parameter, $b_{\rm max, unres}$, is independent of mass resolution and depends only on the gravitational softening and halo scale length, for which we provide a relation in Eq.~(\ref{optimal_eta_scaled}). For most large-volume cosmological simulations and haloes hosting SMBHs, $b_{\rm max, unres}\approx 0.5\, \epsilon_g$. We emphasise that the exact value also depends on the softening spline employed in the simulations. Our results, in this case, are specific to the softening spline used in {\small GADGET-2} \citep{gadget2} and related codes. }

 \item{We also obtain a relation between the number of particles required to accurately compute the local density and velocity dispersion as a function of the number of particles present in the halo (Eq.~\ref{kernel_eq}). We show that this substantially improves the error on these estimated quantities, compared to using a fixed number of neighbouring particles (e.g. 64 particles as in \citealt{tremmel}). Directly counting the number of particles moving slower than the black hole (which are the ones contributing to dynamical friction) is more accurate than applying the Maxwellian approximation, as in \citet{chen}, which underestimates the density near the centres of cuspy Hernquist halos. As a result, the model of \citet{chen} requires $b_{\rm max, unres}\approx 6\, \epsilon_g$ to sink the black holes faster, although, at high resolution, this combination of parameters tends to sink black holes too fast.  }

 \item{We have quantified the effects of stochasticity on black hole orbits by varying the background halo mass resolution. We found that even in cases with $M_{\rm BH}/M_{\rm DM} = 100$ stochastic effects can alter a black hole's orbit, although the time it takes for the effects to become noticeable in terms of sinking time are longer. The magnitude of stochastic heating increases as the $M_{\rm BH}/M_{\rm DM}$ ratio and $\epsilon_g$ decrease. We have derived an order-of-magnitude correction for the effects of stochasticity, which, in effect, enhances dynamical friction (Eq.~\ref{stoch_correction}). This correction is, in theory, equivalent to increasing the dynamical mass of the black hole by the dark matter particle mass. We emphasize that, due to the random nature of this problem, the recovered sinking timescales of black holes can become highly inaccurate, particularly for $M_{\rm BH}/M_{\rm DM} \leq 5$. Our correction does, however, guarantee that the black hole sinks. }

 \item{We have validated our model on a number of test cases and levels of resolution, for circular and eccentric orbits. We showed that, for a set of random seeds, we on average accurately reproduce the black hole orbits in {\small KETJU}, with scatter in black hole sinking times increasing as $M_{\rm BH}/M_{\rm DM}$ decreases. }
 
\end{enumerate}

Of course, our model has only been tested in spherical, equilibrium $\rho \propto r^{-1}$ halos and its validity may change when applied to different density distributions (particularly cored halos, see e.g.~\citealt{antonini_merritt}), flattened systems and systems undergoing a merger, which would introduce spatial and velocity anisotropy \citep{penarrubia_flat}. On the other hand, a great advantage of this model and previous models of its kind, is that some dynamical friction is already captured in the simulations, and the correction we apply is only a local one. In other words, the accuracy of the recovered black hole motion should depend only on whether the Chandrasekhar formula represents well the impact of dynamical friction within $\sim \epsilon_g$, which, depending on the spatial resolution of the simulation, may be a very small scale. Moreover, like models of \citet{tremmel} and \citet{chen}, as well as repositioning schemes, our dynamical friction prescription does not conserve momentum. This may lead to unrealistic properties of the particles surrounding the black hole, particularly in regions where the dynamical friction correction is strong. We discuss the implications of this in detail in Appendix C. Ultimately, testing the full range of validity of our model requires more sophisticated numerical setups and cosmological zoom simulations, as well as complementary high-resolution simulations with {\small KETJU}, all of which we will leave to future work. 

Cosmological hydrodynamics simulations are undoubtedly very powerful tools to study black holes and how they affect their host galaxies. Predictions can be made on black hole growth and merger rates, halo occupation distribution and populations of off-centre and wandering black holes. These predictions, however, have many modelling uncertainties. The treatments of black hole seeding, accretion, feedback, mergers, and interactions with other black holes all play an important role. In this work, we focused on the problem of modelling the effects of dynamical friction on black hole dynamics and introduced a subgrid model that has been calibrated on simulations where dynamical friction is fully resolved. We have shown that our model is able to accurately capture black hole sinking timescales when compared to {\small KETJU}, achieving higher performance compared to alternatives in the literature. It is therefore ready to be applied in the next generation of cosmological hydrodynamics simulations.

\section*{Acknowledgements}

 The authors would like to thank Yueying Ni, Christian Partmann and Simon White for helpful discussions. This research was carried out using the High Performance Computing resources of the FREYA cluster at the Max Planck Computing and Data Facility (MPCDF, https://www.mpcdf.mpg.de) in Garching operated by the Max Planck Society (MPG). This project was developed as part of the Simons Collaboration on ``Learning the Universe''. This research benefited from the use of {\sc emcee} \citep{emcee}, {\sc numpy} \citep{numpy}, {\sc scipy} \citep{scipy} and {\sc matplotlib} \citep{matplotlib}.

\section*{Data availability}
The data analysed in this article can be made available upon reasonable request to the corresponding author.




\bibliographystyle{mnras}
\bibliography{bh.bib} 




\appendix

\renewcommand\thefigure{A\arabic{figure}}    
\setcounter{figure}{0} 
\begin{figure}
    \centering
    \includegraphics[width = \columnwidth]{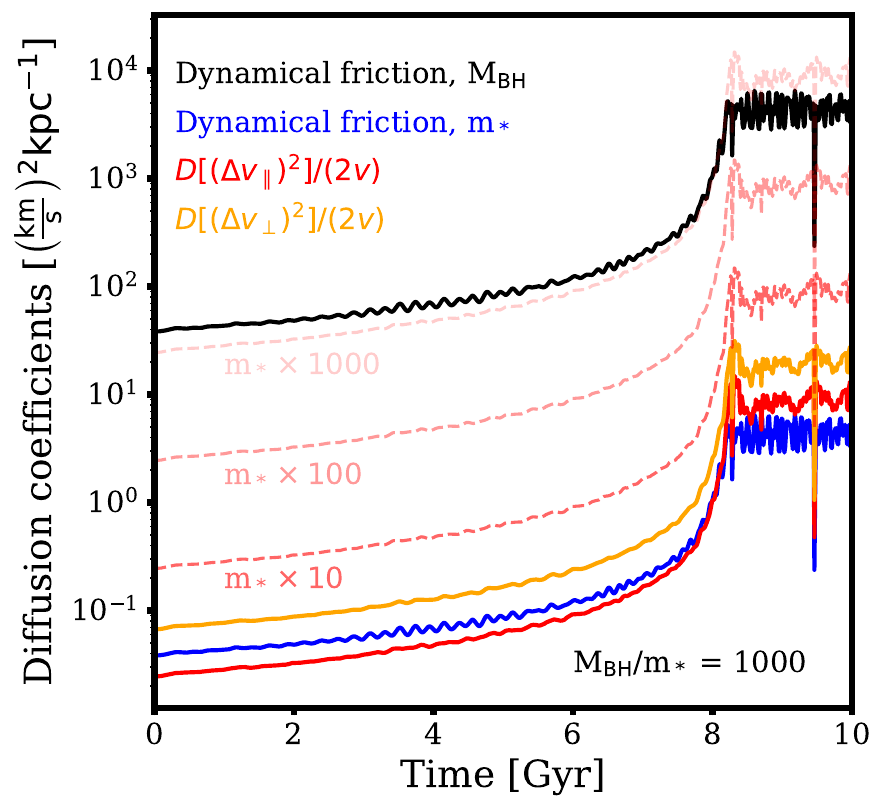}
    \caption{The magnitude of the diffusion coefficients for a $10^9\,{\rm M}_{\odot}$ black hole on a circular orbit inside a $10^{13}\,{\rm M}_{\odot}$ Hernquist halo, with the black hole-to-background particle mass ratio $M_{\rm BH}/m_{\rm *}=1000$. The black line shows the magnitude of dynamical friction due to the mass of the black hole, the blue line shows additional dynamical friction arising from the mass of the background particles, $m_*$. The red line gives the magnitude of the stochastic force in the direction parallel to the direction of motion of the black hole, and the orange line displays the magnitude of the stochastic force perpendicular to the direction of motion. The dashed red lines indicate how the magnitude of the second-order parallel diffusion coefficient would change if the background particle mass increases (and the mass ratio decreases).}
    \label{fig_a1}
\end{figure}

\section{The diffusion coefficients for a Hernquist halo}

In Fig.~\ref{fig_a1}, we demonstrate how the dynamical friction forces compare to the stochastic forces for our fiducial Hernquist halo, with $M_{\rm BH}/M_{\rm DM}=1000$. The diffusion coefficients were computed through explicitly evaluating Eqns.~(\ref{cooling_eq}), (\ref{heating_eq}) and (\ref{heating_eq2}), assuming no softening. It can be seen that the dynamical friction force due to the background particle mass (blue line) is of similar order of magnitude as the second-order diffusion coefficients (red and orange lines), which justifies our approximation of the stochastic force in Eqn.~(\ref{stoch_correction}). Note that the second-order diffusion coefficient in the perpendicular direction, $D[(\Delta v_{\perp})^2]/(2v)$, corresponds to 4 possible directions in which the stochastic force can act, while the parallel second-order diffusion coefficient, $D[(\Delta v_{\parallel})^2]/(2v)$, acts only in one direction (the direction of motion of the black hole).

The dashed red lines show how $D[(\Delta v_{\parallel})^2]/(2v)$ would change if the mass of the background particles were to increase (reducing $M_{\rm BH}/M_{\rm DM}$), increasing the impact of stochasticity relative to that of dynamical friction.

\section{Maximum unresolved impact parameter in the Tremmel model}

\renewcommand\thefigure{A\arabic{figure}}  
\setcounter{figure}{1} 
\begin{figure*}
    \centering
    \includegraphics[width = 2\columnwidth]{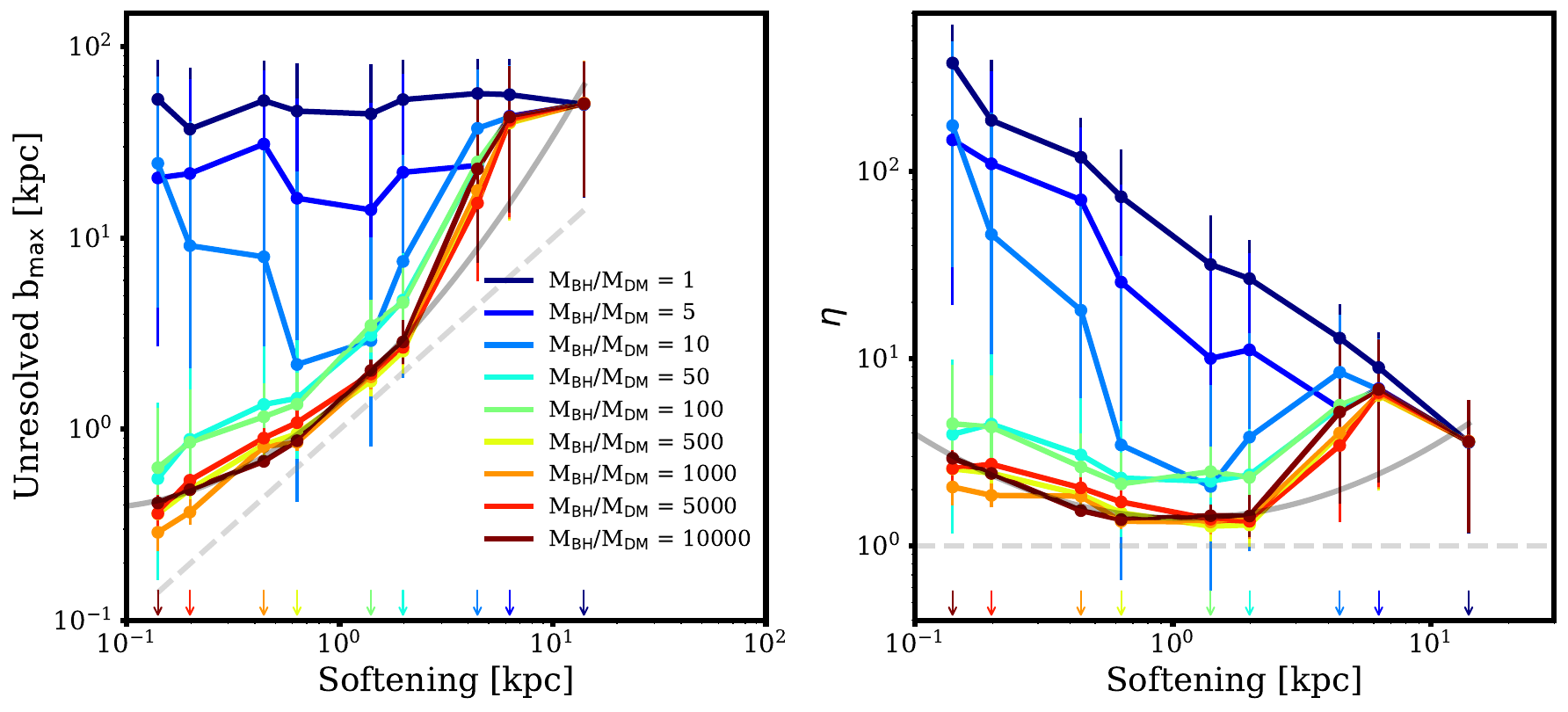}
    \caption{As Fig.~\ref{fig6}, but using the \citet{tremmel} definition of $b_{\rm min}$.}
    \label{fig_a2}
\end{figure*}

As in Section~\ref{themethod}, we here derive the optimal value of $\eta$ for the case $b_{\rm min} = GM_{\rm BH}/v_{\rm BH}^2$. This is shown in Fig.~\ref{fig_a2}, resulting in a relation
\begin{equation}
\eta_{\rm T15} = \rm min[ 1.41~(\epsilon_g/{\rm kpc})^{-0.04}~10^{~0.08 \ln^2(\epsilon_g/{\rm kpc})},~3.0 ].
\end{equation}
Again, we emphasise that this relation is only applicable for the softening spline used in {\small GADGET} and {\small AREPO}.

Interestingly, the relation is clearly offset for the cases $M_{\rm BH}/M_{\rm DM}=50$ and $100$, unlike what we obtained under a different definition of $b_{\rm min}$, suggesting different values of $\eta$ are required as the resolution changes. This is reflected in the left panel of Fig.~\ref{fig2}, where the simulation with $M_{\rm BH}/M_{\rm DM}=50$ requires stronger dynamical friction to reduce the sinking time. In the simulation with $M_{\rm BH}/M_{\rm DM}=100$, the black hole sinking is also delayed early on, though it subsequently catches up at higher resolution levels due to the increased ellipticity of the orbit, induced by noise. As the local velocity dispersion appears to be an important component of $b_{\rm min}$, we are able to achieve higher convergence across resolution levels with our method compared to \citet{tremmel}.

\section{Momentum conservation}

\renewcommand\thefigure{A\arabic{figure}}  
\setcounter{figure}{2} 
\begin{figure*}
    \centering
    \includegraphics[width = 2\columnwidth]{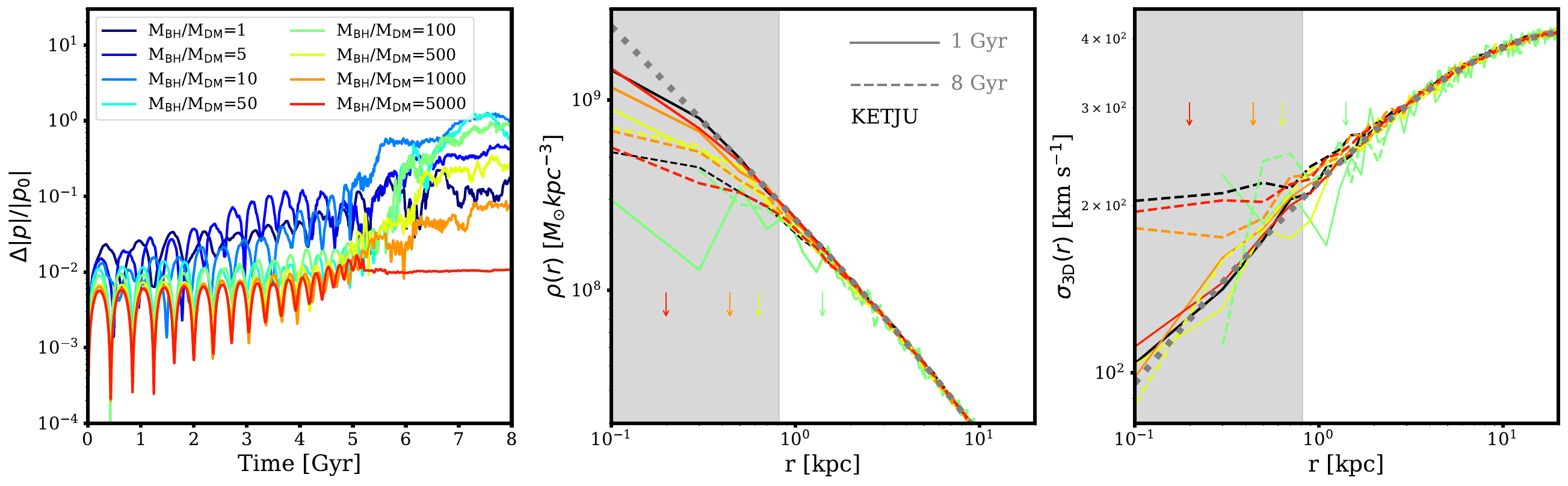}
    \caption{\textit{Left:} fractional change in the magnitude of the total momentum of the simulated system (a $10^{13}\,{\rm M}_{\odot}$ Hernquist halo with a $10^{9}\,{\rm M}_{\odot}$ black hole) when our dynamical friction model is applied. Different colours correspond to different background particle resolution and softening (see Table~\ref{table1}). The black holes sink at $\sim$5 Gyr. \textit{Middle:} the change in the density profile after 1~Gyr (solid line) and 8~Gyr of evolution (dashed line) for the 4 highest-resolution cases from the left panel. Black lines show the results of the simulation with {\small KETJU} (where momentum is essentially conserved, with $\Delta|p|/|p_0| \sim 10^{-8}$). The grey dotted line is the analytical Hernquist density profile. The grey shaded band covers the region of the halo containing $10^9\,{\rm M}_{\odot}$ of dark matter in the initial conditions, equal to the black hole particle mass. The arrows show the gravitational softening used in each simulation. \textit{Right:} as the middle panel, but showing the change in 3-dimensional velocity dispersion.}
    \label{fig_a3}
\end{figure*}

Similarly to the models of \citet{tremmel} and \citet{chen}, as well as repositioning schemes, our dynamical friction model does not conserve the momentum of the simulated system. This is demonstrated on the left panel of Fig.~\ref{fig_a3}, where we show the fractional change of the magnitude of momentum compared to the initial conditions. It is evident that in the lower-resolution simulations (where the softening is large), the discrepancy is more significant and the change in momentum can reach $\sim10$~per~cent of the initial value by the time the black hole sinks (at $\sim 5$~Gyr). After the black hole reaches the centre of the halo, the change in the momentum of the system can become extremely high, especially at low resolution. This is because the black hole is not fully stable at the halo centre and, as long as $b_{\rm min}$ is smaller than $b_{\rm max, unres}=\eta \epsilon_g$, the deceleration of the black hole in high-density regions computed with our model is extremely large. For the high-resolution case, $M_{\rm BH}/M_{\rm DM}=5000$ and $\epsilon_g = 0.2$~kpc, since the central regions are well resolved and $b_{\rm min}$ becomes greater than $b_{\rm max, unres}$ (setting the dynamical friction correction to 0) the black hole is able to stay at the centre of the halo and the momentum thereafter is essentially conserved. 

As the black hole (or any massive infalling object) is slowed down due to dynamical friction, it loses angular momentum, while the surrounding particles would, in turn, gain momentum. This process has previously been invoked as a mechanism for transforming cold dark matter cusps in the centres of haloes into constant-density cores \citep{el_zant,elzant2, cole_core_df}. In the middle panel of Fig.~\ref{fig_a3}, we demonstrate that our simulation with {\small KETJU} of a $10^{9}\,{\rm M}_{\odot}$ black hole sinking to the centre of a $10^{13}\,{\rm M}_{\odot}$ Hernquist halo (black solid and dashed lines) also predicts a shallower dark matter density profile once the black hole has sunk to the halo centre. An increase in the velocity dispersion is also evident below a radius of 800~pc (where the initial Hernquist profile contains $10^9\,{\rm M}_{\odot}$ in dark matter). In coloured lines, we show how these halo properties differ when our dynamical friction model is applied. When the softening is small (< 400~pc), the flattening of the density profile is partially recovered (as momentum exchange still occurs on the scales where dynamical friction is resolved) and the increase in the velocity dispersion of the surrounding particles can also be seen as the surrounding particles are heated by the black hole. When the softening is large (> 400~pc), the slope of the density profile does not substantially change and neither does the velocity dispersion. Although the lower-resolution cases do have a central density core, this is primarily induced by the gravitational softening \citep{barnes_soft}, while the background particle kinematics remain cold.

In practice, the lack of momentum conservation would lead to underestimated velocity dispersion of background stars and dark matter in the vicinity of the black hole, with the effect being strongest when the gravitational softening is large, for very massive black holes and in dense regions, where the dynamical friction is strong. To circumvent this issue, one could, in principle, artificially inject momentum into the surrounding particles such that total momentum is manifestly conserved. It is, however, unclear how to distribute this momentum, such that the resultant dynamics of the surrounding particles faithfully reflect momentum transfer in dynamical friction. Perhaps, a weighting scheme for particles with distance $d<\eta \epsilon_g$ of the black hole that follows Eqn.~(9) of \citet{ma_hopkins} could be used, though this will likely break down at low resolution if the black hole is not stable at the centre of the halo (see the left panel of Fig.~\ref{fig_a3}), leading to an unnecessarily large cumulative momentum injection into the surrounding matter. For now, it is clear that spatial and kinematic properties of particles surrounding a massive black hole in galaxy centres cannot be taken at face value in simulations where momentum exchange due to dynamical friction is not fully accounted for.

\bsp	
\label{lastpage}
\end{document}